\documentstyle[preprint,aps,eqsecnum,floats,graphicx]{revtex}

\tighten  
\def\beq{\begin{equation}}
\def\eeq{\end{equation}}

\def\beqn{\begin{eqnarray}}
\def\eeqn{\end{eqnarray}}
\def\nn{\nonumber\\ }

\def\A{{\mathcal A}}   
\def\B{\hat B}         
\def\P{\hat P}
\def\b{\hat b}

\def\F{{\mathcal F}}   
\def\L{{\mathcal L}}   
\def\O{{\mathcal O}}

\def\Im{{\mbox{\,Im\,}}}

\def\vecr{\bbox{r}}
\def\vecq{\bbox{q}}
\def\vecp{\bbox{p}}
\def\vecj{\bbox{j}}
\def\vecP{\bbox{P}}
\def\vecR{\bbox{R}}
\def\vecE{\bbox{E}}
\def\vecB{\bbox{B}}
\def\vecD{\bbox{D}}
\def\vecepsilon{\bbox{\epsilon}}
\def\vecnabla{\bbox{\nabla}}
\def\vecrho{\bbox{\rho}}
\def\vectau{\bbox{\tau}}
\def\vecpi{\bbox{\pi}}

\def\vecMag{\bbox{\mathcal M}}   
\def\vecPol{\bbox{\mathcal P}}
\def\Pol{{\mathcal P}}

\begin{document}
\def\thefootnote{\arabic{footnote}}

\preprint{MKPH-T-01-02; ETC$^*$-01-002}

\title{Generalized dipole polarizabilities \\
   and the spatial structure of hadrons}

\author{
A.I. L'vov,$^1$
S. Scherer,$^2$
B. Pasquini,$^3$
C. Unkmeir,$^2$
D. Drechsel$^2$
}

\address{
$^1$P.N. Lebedev Physical Institute, Leninsky Prospect 53,
      Moscow, 117924, Russia
\\[1ex]
$^2$Institut f\"ur Kernphysik, Johannes Gutenberg-Universit\"at,
      J.J. Becher-Weg 45, D-55099 Mainz, Germany
\\[1ex]
$^3$ECT$^\ast$, European Centre for Theoretical Studies in Nuclear
      Physics and Related Areas,\\
      I-38050 Villazzano (Trento) and INFN, Trento (Italy)
}

\maketitle

\begin{abstract}

   We present a phenomenological discussion of spin-independent, generalized
dipole polarizabilities of hadrons entering the virtual Compton scattering
process $\gamma^\ast h\to \gamma h$.
   We introduce a new method of obtaining a tensor basis with appropriate
Lorentz-invariant amplitudes which are free from kinematical
singularities and constraints.
   The result is summarized in terms of a compact effective Lagrangian.
   We then motivate a gauge-invariant separation into a generalized Born
term containing ground-state properties only, and a residual contribution
describing the model-dependent internal structure.
   The generalized dipole polarizabilities are defined in terms of
Lorentz-invariant residual amplitudes.
   Particular emphasis is laid on a physical interpretation of these
quantities as characterizing the spatial distributions
of the induced electric polarization and magnetization of hadrons.
   It is argued that three dipole polarizabilities, namely the longitudinal
electric $\alpha_L(q^2)$, the transverse electric $\alpha_T(q^2)$, and the
magnetic $\beta(q^2)$ ones are required in order to fully reconstruct local
polarizations induced by soft external fields in a hadron.
   One of these polarizabilities, $\alpha_T(q^2)$, describes an effect of
higher order in the soft final-photon momentum $q'$.
   We argue that the associated spatial distributions obtained via the
Fourier transforms in the Breit frame are
meaningful even for such a light particle as the pion.
   The spatial distributions are determined at large distances
$r \sim 1/m_\pi$ for pions, kaons, and octet
baryons by use of ChPT.

\end{abstract}



\section{Introduction}

Recently,  Compton scattering with virtual photons (virtual Compton
scattering, or VCS) has attracted considerable experimental and
theoretical interest (for an overview see, e.g.,
\cite{guichon98,scherer99,vanderhaeghen00} and references therein).
Both the near-threshold \cite{guichon95} and
high-momentum transfer regimes \cite{kroll96,radyushkin97,ji97} of VCS
turn out to be very informative for studying the structure of hadrons.

   In the below-threshold region a set of structure functions, the
so-called generalized polarizabilities (GPs), has been introduced
in Ref.\ \cite{guichon95} in order to parametrize structure-dependent effects
in the VCS amplitude to leading order in the final-photon momentum $q'$.
   These GPs contain model-dependent information beyond the low-energy
theorem (LET) of real \cite{Low_1954,GellMann_1954}
and virtual Compton scattering \cite{Scherer_96}.
   The first experimental results for generalized polarizabilities of the
proton have recently been obtained at the Mainz Microtron (MAMI) for a
four-momentum squared of $Q^2=0.33$ GeV$^2$ \cite{roche00}.
   Additional experiments aiming at proton polarizabilities
are presently carried out at Jefferson Lab
\cite{Audit_1993} and MIT-Bates \cite{Shaw_1997}.
   A sensitivity study of inelastic high-energy pion-electron scattering
on the pion VCS amplitude has been performed as part of the Fermilab
E781 SELEX experiment \cite{moinester99,Ocherashvili00}.

   Theoretical predictions of the nucleon GPs have been obtained in different
frameworks such as the constituent quark model
\cite{guichon95,liu96,pasquini98,pasquini00a},
the linear $\sigma$-model \cite{metz96},
chiral perturbation theory (ChPT) \cite{hemmert97,hemmert00},
the Skyrme model \cite{kim97},
models with effective photon-pion-nucleon Lagrangians
\cite{vanderhaeghen96,korchin98}, and
dispersion theories \cite{pasquini00b,pasquini01}.
   Pion and kaon GPs have been discussed at the one-loop level of
ChPT \cite{unkmeir00,fuchs00},

   In the present paper we are not so much concerned with specific results for
GPs as obtained from various models of hadrons.
   Our investigations are rather directed to an analysis of general properties
of GPs as well as their physical interpretation which, in our opinion, was
only insufficiently covered previously.
   We mainly consider kinematical aspects, relegating
dynamical aspects of properties of the generalized polarizabilities such as
sum rules, dispersion relations, etc.\ to a future publication.
   In order to avoid complications related with the spin of the target
   we only discuss the simplest case of a (pseudo) scalar particle like the
pion.
   Our considerations may as well be applied to the spin-averaged part of the
VCS amplitude for a target of arbitrary spin.
   For illustrative purposes we make use of results obtained in the framework
of ChPT.

   Our work is organized as follows.
   In Sec.\ II, we introduce a new method of obtaining invariant amplitudes of
electromagnetic reactions free from kinematical singularities and
constraints,  using (double) virtual Compton scattering as an
example.
   Section III contains a motivation for choosing a specific (though standard)
form of the Born amplitude which is used for separating convectional
from internal contributions in the scattering amplitude.
   In Sec.\ IV, we show how generalized dipole polarizabilities can be defined
in a Lorentz-invariant manner from the invariant amplitudes without
introducing inconvenient singular kinematical factors.
   In particular, we show that it is natural to introduce one more dipole
polarizability, namely, the transverse electric one,
which is needed in order to recover the electric polarization of the target
induced by a soft external electric field.
   This quantity does not contribute in the $\O(q')$ limit of previous
analyses.
      We provide another explanation why to this order
only two of the three spin-independent polarizabilities are accessible in
photon electroproduction experiments.
   We give a physical interpretation of the generalized dipole polarizabilities
in terms of densities of the induced electric and magnetic polarizations.
   In Sec.\ V we argue that the associated space distributions are meaningful
even for a light particle like the pion.
   We determine such space distributions at large distances $r \sim 1/m_\pi$
derived from various form factors using the predictions of ChPT.
   In passing, we also include SU(3)$_f$ extensions of previous results
obtained within two-flavor HBChPT.
   Analytical results of spatial distributions obtained via dispersion
relations are relegated to the Appendix.

\section{Tensor basis and invariant amplitudes of VCS}

    In our discussion of virtual Compton scattering off a spinless hadron,
we will often refer to this hadron as a ``pion,'' although our general
considerations also hold true for other spinless particles, nuclei, and even
atoms, as well as for spin-averaged amplitudes.
   We start our analysis with an investigation of the general
kinematical structure of the pion VCS amplitude $T_{\rm VCS}$
for the case of two virtual photons, assuming that both initial and final
pions are on shell.
   Our aim is to construct a tensor basis and a related
set of Lorentz-invariant amplitudes $B_i$,
that provides a complete parametrization of $T_{\rm VCS}$.
   We require all $B_i$ to be free from kinematical
singularities and constraints,
because this simplifies the classification
of low-energy characteristics of the pion, and also
provides technical advantages, for instance when discussing
dispersion relations.

   The problem of finding a set of amplitudes for VCS
was already addressed and solved by Tarrach for both spin-0
and spin-1/2 targets \cite{tarrach75} by using a projection
technique originally proposed by Bardeen and Tung \cite{bardeen68}.
   Although, in principle, we could directly use the results of
Ref.\ \cite{tarrach75}, we prefer to construct the tensor basis
again, first, in order to demonstrate a very simple and powerful
alternative method which avoids projections and, second, in order
to rearrange the VCS tensor in a manifestly gauge-invariant form
and to split it into structures contributing to real Compton scattering,
VCS with one photon virtual and, finally, VCS with both photons virtual.

   To begin with, we define the amplitude $T_{\rm VCS}$ of
virtual Compton scattering,
\beq
\label{VCSreaction}
   \gamma(\epsilon,q) + \pi(p)
        \to \gamma(\epsilon',q') + \pi(p'),
\eeq
as
\beq
  T_{\rm VCS} = \epsilon_\mu \epsilon_\nu^{\prime *} T^{\mu\nu},
\eeq
where $T_{\mu\nu}$ is the Compton tensor given in terms of the
covariant (Wick) $T_W$-product of the electromagnetic currents,
\beq
     T_{\mu\nu} = \int \langle\pi(p')|
        iT_W[j_\mu(x) j_\nu(0)] |\pi(p)\rangle  \,e^{-iq\cdot x}\,d^4x.
\label{T*}
\eeq
We normalize all single-particle states as
\beq
\label{state-norm}
     \langle \vecp' | \vecp \rangle
   = 2p_0 (2\pi)^3 \delta^3(\vecp - \vecp'),
\eeq
so that in case of real photons the $S$-matrix reads
\beq
   S_{fi} = i (2\pi)^4 \delta^4(p+q-p'-q') \,T_{\rm VCS}.
\label{S_fi}
\eeq
  Due to four-momentum conservation, $p+q=p'+q'$, the tensor
$T_{\mu\nu}$ depends on three linearly independent vectors
$P$, $Q$, and $R$:
\beq
\label{pqr}
    P=\frac12(p+p'),\quad
    Q=\frac12(q+q'),\quad
    R=\frac12(p'-p) = \frac12(q-q').
\eeq
   Since we only consider the case of initial and
final pions on the mass shell, the vectors $P$, $Q$, and $R$
are constrained by
\beq
  P^2 = M^2 - R^2,  \quad P\cdot R=0,
\eeq
where $M$ is the pion mass.
   Hence, we can choose four independent kinematical invariants
which, for the moment, we take to be $Q^2$, $R^2$,
$P\cdot Q$, and $Q\cdot R$.

   The discrete symmetries as well as gauge invariance impose
restrictions on the general form of the Compton tensor.
   For charged pions the combination of pion crossing with
charge-conjugation symmetry results in%
\footnote{For the neutral pion (but not for the neutral kaon),
which is its own antiparticle, only pion crossing
is required to obtain Eq.\ (\ref{crossing1}).}
\beq
\label{crossing1}
   T_{\mu\nu}(P,Q,R) = T_{\mu\nu}(-P,Q,R),
\eeq
   whereas photon crossing yields
\beq
\label{crossing2}
   T_{\mu\nu}(P,Q,R) = T_{\nu\mu}(P,-Q,R).
\eeq
   Gauge invariance requires
\beqn
\label{gauge-invariance}
   q _\mu T^{\mu\nu} & = & (Q+R)_\mu T^{\mu\nu} =0, \nn
   q'_\nu T^{\mu\nu} & = & (Q-R)_\nu T^{\mu\nu} =0,
\eeqn
   where the first and second equation are not independent
once photon-crossing symmetry is imposed.
   Finding a solution to Eqs.\ (\ref{gauge-invariance}) without
introducing kinematical singularities or
constraints
is the main challenge in constructing appropriate amplitudes.

   Because of parity conservation the Compton tensor transforms
as a proper second-rank Lorentz tensor.
   The most general $T_{\mu\nu}$ satisfying the
crossing-symmetry conditions of Eqs.\ (\ref{crossing1})
and (\ref{crossing2}) can be written as
a linear combination
\beq
\label{T-gen-A}
  T_{\mu\nu} = \sum_{i=1}^{10} \tau^i_{\mu\nu} A_i
\eeq
of ten basis tensors $\tau^i_{\mu\nu}$
which include the metric tensor $g_{\mu\nu}$ and 9 bi-linear
products of $P$, $K$, and $Q$:\footnote{Due to parity, structures
containing a single fully antisymmetric tensor
$\epsilon_{\mu\nu\alpha\beta}$ are excluded.}
\beqn
\label{basis-tau}
  \tau^1_{\mu\nu}     &=&  g_{\mu\nu},  \nn
  \tau^2_{\mu\nu}     &=&  P_\mu P_\nu, \nn
  \tau^3_{\mu\nu}     &=&  Q_\mu Q_\nu, \nn
  \tau^4_{\mu\nu}     &=&  R_\mu R_\nu, \nn
  \tau^5_{\mu\nu}     &=& (P_\mu Q_\nu + P_\nu Q_\mu)(P\cdot Q),       \nn
  \tau^6_{\mu\nu}     &=& (P_\mu Q_\nu - P_\nu Q_\mu)(P\cdot Q)(Q\cdot R), \nn
  \tau^7_{\mu\nu}     &=& (P_\mu R_\nu + P_\nu R_\mu)(P\cdot Q)(Q\cdot R), \nn
  \tau^8_{\mu\nu}     &=& (P_\mu R_\nu - P_\nu R_\mu)(P\cdot Q),       \nn
  \tau^9_{\mu\nu}     &=& (Q_\mu R_\nu + Q_\nu R_\mu)(Q\cdot R),       \nn
  \tau^{10}_{\mu\nu}  &=& (Q_\mu R_\nu - Q_\nu R_\mu).
\eeqn
   Appropriate factors of $P\cdot Q$ have been introduced such
that all $\tau^i_{\mu\nu}$ are even functions of $P$, as
required by Eq.~(\ref{crossing1}).
   Correspondingly, factors of $Q\cdot R$ provide for
photon-crossing-even basis tensors [see Eq.\ (\ref{crossing2})].
   With such a choice of $\tau^i_{\mu\nu}$, all the functions
$A_i$ depend on the crossing-even variables $Q^2$, $R^2$,
$(P\cdot Q)^2$, and $(Q\cdot R)^2$.

   At this point it might be worthwhile to explain why introducing
factors of $P\cdot Q$ and $Q\cdot R$ into
Eqs.\ (\ref{basis-tau}) is harmless to the analytical properties
of the functions $A_i$ in Eq.\ (\ref{T-gen-A}).
   For that purpose, let us omit in Eqs.\ (\ref{basis-tau}) all
factors of $P\cdot Q$ or $Q\cdot R$, and denote the resulting
basis tensors by $\overcirc\tau^i_{\mu\nu}$ with $\overcirc A_i$
the corresponding functions of the
expansion of $T_{\mu\nu}$ in terms of the ``reduced''
tensors $\overcirc\tau^i_{\mu\nu}$.
   Let us further assume that the tensor $T_{\mu\nu}$ results from
some set of Feynman diagrams consistent with all symmetries.
   Consider an arbitrary Feynman diagram denoted by $G$.
   Obviously, its contribution to $T_{\mu\nu}$ can be expressed
in terms of $g_{\mu\nu}$ or bi-linear products of
four-momenta (these are just the $\overcirc\tau^i_{\mu\nu}$)
multiplied by scalar coefficients.
   Any such coefficient will at most have {\em dynamical}
singularities related with propagators of intermediate particles
but no {\em kinematical} singularities.
   Stated differently, all individual contributions
$\overcirc A_i^G$ to the functions $\overcirc A_i$ are free
from kinematical singularities.
   In general, a single contribution is not separately
crossing symmetric.
   Eventually, crossing symmetry of $T_{\mu\nu}$ is obtained after adding
one or several crossed partners, $G_c$, of the diagram $G$.
   Let us consider, for instance, the tensor structure
$\overcirc\tau^5_{\mu\nu} = P_\mu Q_\nu + P_\nu Q_\mu$
which is odd under $P\to -P$.
   We can then write
\beq
\label{A_5ex}
   \overcirc A_5^{G+G_c}(x) = \overcirc A_5^G(x) - \overcirc A_5^G(-x),
\eeq
where we introduced $x=P\cdot Q$ and, for brevity, omitted
$P$-independent arguments like $Q^2$.
   The second term in Eq.\ (\ref{A_5ex}) represents the
contribution of the crossed diagram $G_c$ which makes the amplitude
$\overcirc A_5^{G+G_c}(x)$ odd.
   Note that both contributions are non-singular as
$x\to 0$.
   From Eq.\ (\ref{A_5ex}) we conclude that
$\overcirc A_5^{G+G_c}(x) = x A_5^{G+G_c}(x)$,
where $A_5^{G+G_c}(x)$ is an even function of $x$
which has no pole at $x=0$ and which therefore has no
kinematical singularities at all.
   In other words, the sum of all Feynman diagrams
contains the term $\tau^5_{\mu\nu} A_5$, the tensor
$\tau^5_{\mu\nu}$ carrying a factor of $P\cdot Q$ and
the function $A_5$ having no kinematical singularities.

   The same consideration immediately applies to the tensors
$\tau^8_{\mu\nu}$ and $\tau^9_{\mu\nu}$.
In the case of $\tau^6_{\mu\nu}$ or $\tau^7_{\mu\nu}$,
we have to apply the procedure given by Eq.\ (\ref{A_5ex}) twice ---
first for showing that the corresponding function $\overcirc A_i$
contains the factor of $P\cdot Q$,
and second for showing that it contains the factor of
$Q\cdot R$ as well. Eventually, we conclude that all
the coefficients $A_i$ in Eq.\ (\ref{T-gen-A}) are free from
kinematical singularities.

   By means of the factors $P\cdot Q$ and $Q\cdot R$
in the tensor basis Eq.\ (\ref{basis-tau}) we can
solve the constraints of Eqs.\ (\ref{gauge-invariance})
without introducing singular coefficients and without
using the projectors suggested in Ref.\ \cite{bardeen68}.
   Indeed, inserting Eq.\ (\ref{T-gen-A}) for
$T^{\mu\nu}$ into Eqs.\ (\ref{gauge-invariance}) and collecting
coefficients of the independent four-momenta $P$, $Q$, and $R$,
we obtain a set of six linear equations in the functions $A_i$,
of which only five are independent:
\beqn
   && A_2 + Q^2 A_5 - (Q\cdot R)^2 A_6 + (Q\cdot R)^2 A_7 - R^2 A_8 =0, \nn
   && A_5 - Q^2 A_6 + R^2 A_7 - A_8 =0, \nn
   && A_1 + Q^2 A_3 + (P\cdot Q)^2 A_5 + (Q\cdot R)^2 A_9 - R^2 A_{10} =0, \nn
   && A_3 + (P\cdot Q)^2 A_6 + R^2 A_9 - A_{10} =0, \nn
   && A_1 + R^2 A_4 + (P\cdot Q)^2 A_8 + (Q\cdot R)^2 A_9 + Q^2 A_{10} =0, \nn
   && A_4 + (P\cdot Q)^2 A_7 + Q^2 A_9 + A_{10} =0.
\eeqn
   One can now express $A_i$, $i=1,\cdots,5$, in terms of the
remaining functions $A_i$, $i=6,\cdots,10$, without singular
coefficients:
\beqn
\label{a12345}
  && A_1= R^2 (P\cdot Q)^2 A_7 - (P\cdot Q)^2 A_8
    + [Q^2 R^2 -(Q\cdot R)^2] A_9 + (R^2-Q^2) A_{10},  \nn
  && A_2= [(Q\cdot R)^2-(Q^2)^2] A_6
        + [Q^2 R^2-(Q\cdot R)^2] A_7 + (R^2-Q^2) A_8,  \nn
  && A_3= -(P\cdot Q)^2 A_6 - R^2 A_9 + A_{10}, \nn
  && A_4= -(P\cdot Q)^2 A_7 - Q^2 A_9 - A_{10}, \nn
  && A_5=  Q^2 A_6 - R^2 A_7 + A_8.
\eeqn
Using Eqs.\ (\ref{a12345}), we can finally rewrite
Eq.~(\ref{T-gen-A}) as
\beq
  T_{\mu\nu} = \sum_{i=6}^{10} T^i_{\mu\nu}
          A_i(Q^2,R^2,(P\cdot Q)^2,(Q\cdot R)^2),
\eeq
where
\beqn
\label{basis}
 && T^6_{\mu\nu} = [(Q\cdot R)^2-(Q^2)^2] P_\mu P_\nu
          - (P\cdot Q)^2 Q_\mu Q_\nu
\nn && \qquad {}
      + Q^2 (P\cdot Q)(P_\mu Q_\nu + P_\nu Q_\mu)
      + (P\cdot Q)(Q\cdot R)(P_\mu Q_\nu - P_\nu Q_\mu), \nn
 && T^7_{\mu\nu} = R^2 (P\cdot Q)^2 g_{\mu\nu}
     + [Q^2 R^2-(Q\cdot R)^2] P_\mu P_\nu  - (P\cdot Q)^2 R_\mu R_\nu
\nn && \qquad {}
      - R^2 (P\cdot Q)(P_\mu Q_\nu + P_\nu Q_\mu)
      + (P\cdot Q)(Q\cdot R)(P_\mu R_\nu + P_\nu R_\mu), \nn
 && T^8_{\mu\nu} = -(P\cdot Q)^2 g_{\mu\nu} + [R^2-Q^2] P_\mu P_\nu
\nn && \qquad {}
      + (P\cdot Q)(P_\mu Q_\nu + P_\nu Q_\mu)
      + (P\cdot Q)(P_\mu R_\nu - P_\nu R_\mu), \nn
 && T^9_{\mu\nu} = [Q^2 R^2 - (Q\cdot R)^2] g_{\mu\nu} - R^2 Q_\mu Q_\nu
    - Q^2 R_\mu R_\nu + (Q\cdot R)(Q_\mu R_\nu + Q_\nu R_\mu), \nn
 && T^{10}_{\mu\nu} = (R^2 - Q^2) g_{\mu\nu} + Q_\mu Q_\nu
      - R_\mu R_\nu + (Q_\mu R_\nu - Q_\nu R_\mu)
\eeqn
are five basis tensors which explicitly satisfy the
crossing-symmetry and gauge-invariance conditions of Eqs.\
(\ref{crossing1})--(\ref{gauge-invariance}).
   Respectively, the five functions $A_i$, $i=6,\cdots,10$,
can be considered as invariant amplitudes of virtual Compton
scattering which are free from kinematical singularities and
constraints.

   In passing we note that the same method of
constructing a basis and invariant amplitudes free from kinematical
singularities and constraints also works for the VCS amplitude
of a spin-1/2 target such as the nucleon.

   The tensors $T^8_{\mu\nu}$ and $T^{10}_{\mu\nu}$
have exactly the same form as for real Compton scattering
and can easily be identified with the more common notation
\beqn
 && T^8_{\mu\nu} = -(P\cdot q)(P\cdot q') g_{\mu\nu}
      - (q\cdot q') P_\mu P_\nu
      + (P\cdot q') P_\mu q_\nu + (P\cdot q) P_\nu q'_\mu,
\nn
 && T^{10}_{\mu\nu} = q'_\mu q_\nu - (q\cdot q') g_{\mu\nu}.
\eeqn

   However, also the remaining tensors $T^i_{\mu\nu}$,
$i=6,7,9$, and thus the corresponding functions
$A_i$ contribute to RCS.
   This, clearly, is a drawback of the basis of
Eqs.\ (\ref{basis}) and it would be convenient to have, instead,
a basis such that {\em exclusively} the two tensors
$T^8_{\mu\nu}$ and $T^{10}_{\mu\nu}$ contribute to RCS,
another one appears for the case of one virtual photon
and, finally, the two remaining structures
also contribute to $\gamma^\ast\pi
\to\gamma^\ast\pi$.
   To that end let us introduce gauge-invariant combinations of
photon polarizations and momenta,
\beq
   F_{\mu\nu}  = -i
      (q_\mu \epsilon_\nu - q_\nu \epsilon_\mu ), \quad
   F_{\mu\nu}' =  i
      (q_\mu'\epsilon_\nu^{\prime *} - q_\nu'\epsilon_\mu^{\prime *}).
\eeq
   These second-rank tensors represent the Fourier components
of the electromagnetic field-strength tensor
$\F_{\mu\nu}(x) = \partial_\mu A_\nu(x) - \partial_\nu A_\mu(x)$
associated with plane-wave initial and final photons
described by vector potentials $A_\mu(x) = \epsilon_\mu \exp(-iq\cdot x)$ and
$A_\mu'(x) = \epsilon_\mu^{\prime *} \exp(iq'\cdot x)$, respectively.
   In terms of $F_{\mu\nu}$ and $F_{\mu\nu}'$ it turns out
to be rather straightforward to identify structures contributing
for real or virtual photons.

   For example, the $A_{10}$ contribution to the VCS amplitude
reads
\beq
   \epsilon^\mu \epsilon^{\prime * \nu} T_{\mu\nu}^{10} A_{10}
      = -\frac12 F^{\mu\nu} F_{\mu\nu}' A_{10}
\eeq
   which can be interpreted as the matrix element of the effective
Lagrangian\footnote
{Here, we assume that "pion" and "antipion" are
different particles, such as $\pi^+$ and $\pi^-$ or $K^0$ and $\bar{K}^0$.
    For the case of a charged pion, the field $\phi\equiv\pi^+$ is given in
terms of the Hermitian, Cartesian isospin components $\phi_i$  as
$\phi=(\phi_1 - i\phi_2)/\sqrt{2}$
$(\phi^\dagger\equiv\pi^-=(\phi_1 + i\phi_2)/\sqrt{2})$ and destroys a
$\pi^+$ ($\pi^-$).
   In the case of the neutral pion, we have to take
$\phi=\phi^\dagger=\phi_3$ and replace the factor
of 1/4 in Eq.\ (\ref{lagrangian-A10}) by 1/8.
These trivial changes also apply to other Lagrangians written below.}
\beq
    \L =-\frac14 \A_{10} \F_{\mu\nu}\F^{\mu\nu} \phi^\dagger\phi.
\label{lagrangian-A10}
\eeq
   Here $\A_{10}$ represents a differential operator in
terms of (covariant) derivatives acting on both pion and
electromagnetic fields with Fourier components given by the
function $A_{10}$.
   Similarly, the contribution of the amplitude $A_8$ can be
written as
\beq
   \epsilon^\mu \epsilon^{\prime * \nu} T_{\mu\nu}^8 A_8
      = - (P_\mu F^{\mu\nu})(P^\rho F_{\rho\nu}') A_8
\eeq
which results from the effective interaction Lagrangian
\beq
    \L = -\frac12 \A_8 \F^{\alpha\nu} \F_{\beta\nu}
         \P_\alpha \P^\beta \phi^\dagger \phi,
\eeq
where the action of $\P_\alpha\P_\beta \phi^\dagger \phi$ is defined by
\beqn
   \P_\beta \phi^\dagger\phi&=&
         \frac{i}{2}\phi^\dagger D_\beta\phi
        -\frac{i}{2}(D_\beta \phi)^\dagger \phi,
\nn
   \P_\alpha \P_\beta \phi^\dagger\phi&=&
        -\frac{1}{4}\phi^\dagger D_\alpha D_\beta\phi
        +\frac{1}{4}(D_\alpha\phi)^\dagger D_\beta \phi
        +\frac{1}{4}(D_\beta\phi)^\dagger D_\alpha \phi
        -\frac{1}{4}(D_\alpha D_\beta \phi)^\dagger \phi,
\eeqn
and $D_\alpha\phi=\partial_\alpha\phi+ieZ A_\alpha \phi$ with $e>0$,
$e^2/4\pi\approx 1/137$, and $Ze$ denoting the charge of the particle.

   The tensor structures $T_{\mu\nu}^i$, $i=6,7,9$, involve
higher powers of photon momenta and thus are related with
derivatives of the electromagnetic fields.
   Introducing the four-vectors
\beqn
     -iq^\mu F_{\mu\nu} =
         -q^2 \epsilon_\nu + (q\cdot \epsilon) q_\nu,
\nn
      iq^{\prime \mu} F_{\mu\nu}' =
       -q^{\prime 2} \epsilon_\nu^{\prime *}
       + (q'\cdot \epsilon^{\prime *}) q_\nu',
\label{qfqpfp}
\eeqn
which vanish for real photons,
we obtain two more tensors,
namely $T_{\mu\nu}^9$ and $T_{\mu\nu}^6 - T_{\mu\nu}^7$,
by using the identities
\beq
    (q_\mu F^{\mu\nu}) (q^{\prime \rho} F_{\rho\nu}')  =
   \epsilon^\mu \epsilon^{\prime * \nu}
      [4T_{\mu\nu}^9 + (R^2-Q^2) T_{\mu\nu}^{10}]
\eeq
and
\beqn
\label{PdFPdF}
    (P_\nu q_\mu F^{\mu\nu})
    (P^\sigma q^{\prime \rho} F_{\rho\sigma}')
    &=&
   \epsilon^\mu \epsilon^{\prime * \nu}
      [ -2T_{\mu\nu}^6 + 2T_{\mu\nu}^7
    + (R^2+Q^2) T_{\mu\nu}^8 - (P\cdot Q)^2 T_{\mu\nu}^{10}].
\eeqn
   The remaining linear combination
$T_{\mu\nu}^6 + T_{\mu\nu}^7$ is contained in a product
similar to Eq.\ (\ref{PdFPdF}), however with
photon momenta interchanged in one factor:
\beqn
  &&
    (P_\nu F^{\mu\nu} q'_\mu)
    (P^\sigma q^{\prime \rho} F_{\rho\sigma}') +
    (P^\nu F_{\mu\nu}' q^\mu)
    (P_\sigma q_\rho F^{\rho\sigma})
\nn && \qquad =
   \epsilon^\mu \epsilon^{\prime * \nu}
      ( -2T_{\mu\nu}^6 - 2T_{\mu\nu}^7 - 2R^2 T_{\mu\nu}^8) .
\eeqn
   With the above identities the most general VCS amplitude
can be written in the following manifestly gauge-invariant form:
\beqn
\label{ampl}
    T_{\rm VCS} &=& \frac12 F^{\mu\nu} F_{\mu\nu}' B_1
      + (P_\mu F^{\mu\nu})(P^\rho F_{\rho\nu}') B_2
\nn && {}
    + [(P^\nu q^{\prime \mu} F_{\mu\nu})
       (P^\sigma q^{\prime \rho} F_{\rho\sigma}')
   +  (P^\nu q^\mu F_{\mu\nu})
      (P^\sigma q^\rho F_{\rho\sigma}') ]  B_3
\nn && {}
      + (q_\mu F^{\mu\nu}) (q^{\prime \rho} F_{\rho\nu}') B_4
      + (P^\nu q^\mu F_{\mu\nu})
        (P^\sigma q^{\prime \rho} F_{\rho\sigma}') B_5 .
\eeqn
   Here all the invariant amplitudes $B_i$ are free from
kinematical singularities and constraints,
because the transformation from the basis $T^i_{\mu\nu}$ of
Eq.\ (\ref{basis}) to the basis
of Eq.\ (\ref{ampl}) is non-singular.
   This can also be easily seen from the following relations between
the two sets of amplitudes $A_i$ and $B_i$:
\beqn
   B_1 &=& -A_{10} + \frac{(P\cdot Q)^2}{4} (A_6-A_7)
           +\frac{R^2-Q^2}{4} A_9, \nn
   B_2 &=& -A_8 + \frac{R^2}{2}(A_6+A_7) - \frac{R^2+Q^2}{4}(A_6-A_7), \nn
   B_3 &=& -\frac14 (A_6+A_7), \qquad
   B_4 = \frac14 A_9, \qquad
   B_5 = \frac14 (A_7-A_6) .
\eeqn
   The determinant of the transformation expressing $B_i$ ($i=1,\cdots,
5$) in terms of $A_i$ ($i=6,\cdots,10$) is $1/32\neq 0$, independently of the
values of the kinematical variables.

   In view of the rather compact and transparent structure of
Eq.\ (\ref{ampl}), we will use in the following the
parametrization of $T_{\rm VCS}$ given by the amplitudes $B_i$.
   As seen from Eqs.\ (\ref{qfqpfp}) and (\ref{ampl}), only the
amplitudes $B_1$ and $B_2$ are needed to describe real Compton
scattering, because then
$q^\mu F_{\mu\nu}=q^{\prime \mu} F_{\mu\nu}'=0$.
   When only one photon is virtual, one more amplitude ($B_3$)
contributes.
   All five amplitudes $B_i$ enter, when both photons are
virtual.

   Equation (\ref{ampl}) can be interpreted as the matrix element
of the effective interaction
\beqn
\label{L-gen}
  \L &=& \frac14 [ \B_1  (\F_{\mu\nu})^2
      + 2\B_4 (\partial^\mu \F_{\mu\nu})^2 ] \phi^\dagger\phi
\nn && {}
    + \frac12 [ \B_2 \F^{\alpha\mu} \F_{\beta\mu}
   + \B_5 (\partial_\mu \F^{\alpha\mu})(\partial^\nu \F_{\beta\nu})
   -  2\B_3 \F^{\alpha\mu} (\partial_\mu\partial^\nu \F_{\beta\nu}) ]
       \P_\alpha \P^\beta \phi^\dagger \phi,
\eeqn
where $\B_i$ are differential operators acting on all the
fields and determined by their Fourier components $B_i$.

   Of course, after substituting $q\to q_1$, $q'\to -q_2$, $p\to -p_1$, and
$p'\to p_2$, Eq.~(\ref{ampl}) also describes the
general kinematical structure of the amplitude of the crossed
reaction $\gamma(q_1)\gamma(q_2) \to \pi(p_1)\pi(p_2)$ for
on-shell pions.
   Exactly the same considerations apply to any other
spin-0 hadron assuming the same symmetry principles
(Lorentz and gauge invariance, P, T, and C conservation)
and are also applicable to a properly spin-averaged VCS amplitude
for hadrons with finite spin.

   As mentioned before, the functions $B_i$ depend on the four invariants
$Q^2$, $R^2$, $(Q\cdot R)^2$ and $(P\cdot Q)^2$.
   As an alternative, the following combinations of the first 3 quantities
can be used as independent arguments of $B_i$:
$q^2+{q'}^2 = 2(R^2+Q^2)$,
$q\cdot q' = R^2-Q^2$, and $q^2 {q'}^2 = (R^2+Q^2)^2 - 4(Q\cdot R)^2$.
   Thus, we may write
\beq
  B_i = B_i(\nu^2, q\cdot q', q^2 + q^{\prime 2}, q^2 q^{\prime 2}),
\eeq
where $\nu$ is defined as
\beq
   M\nu = P\cdot Q = P\cdot q = P\cdot q'.
\eeq
   Besides being manifestly crossing symmetric, this form of $B_i$ has
the advantage of having a simple limit if one or both photons
become real.

   Finally, the Mandelstam invariants of the VCS reaction
read
\beqn
\label{mandelstam}
   s &=& (p+q)^2  = M^2 + 2M\nu + q\cdot q', \nn
   u &=& (p-q')^2 = M^2 - 2M\nu + q\cdot q', \nn
   t &=& (q-q')^2 = q^2 + q^{\prime 2} - 2q\cdot q'.
\eeqn

\section{The Born terms and gauge invariance}
\label{sec:born}

   In order to describe the internal structure of the pion in terms of its
generalized polarizabilities as tested in virtual Compton scattering,
we first have to isolate a convection contribution which originates in two
successive interactions of the photons with the electromagnetic current
of the pion, resulting in singularities at zero photon momenta.
   For a point-like (pseudo-) scalar particle of electric charge $eZ$,
the interaction with an external electromagnetic field can be described
in terms of the Lagrangian\footnote{Since we do not treat the electromagnetic
field as a dynamical variable, it will not be included in the list
of arguments of the Lagrangian.}
\beq
\label{L-pointlike}
   \L_0(D_\mu\phi,\phi)
   = D_\mu \phi (D^\mu \phi)^\dagger-M^2 \phi\phi^\dagger,
\eeq
   where the covariant derivative
\beq
\label{covder}
    D_\mu \phi = (\partial_\mu + ieZ A_\mu)\phi
\eeq
ensures the invariance of the Lagrangian under the {\em canonical}
gauge transformation
\beq
\label{gauge}
   A_\mu(x) \mapsto A_\mu(x) + \partial_\mu \Lambda(x), \qquad
   \phi(x) \mapsto \exp[-ieZ \Lambda(x)]\phi(x).
\eeq
   However, such a description is not sufficient for an extended particle
and we have to modify the above effective Lagrangian.

   To that end, let us first consider a classical system of $n$ constituents
with charges $e_a=eZ_a$ and masses $m_a$ which is exposed to a static external
potential
$A_0(\vecr)$.
   The electrostatic energy of such a system is given by
\beq
   W=  \sum_{a=1}^n e_a A_0(\vecr_a) = \sum_{a=1}^n e_a \Big[ A_0(\vecR) +
   \rho_{ai}\nabla_i A_0(\vecR) +
   \frac12 \rho_{ai}\rho_{aj}\nabla_i\nabla_j A_0(\vecR) + \cdots \Big],
\eeq
where $\vecR$ is the center of mass of the charge distribution and
$\vecrho_a =  \vecr_a - \vecR$ are relative coordinates.
   In the continuum limit, the expression for a spherically symmetric
distribution reads
\beq
\label{w2}
   W= eZ A_0(\vecR) + \frac{e}{6} \langle Zr_E^2 \rangle
    \vecnabla^2 A_0(\vecR) + \cdots
   =  e F(\vecnabla^2) A_0(\vecR),
\eeq
where $\sum_a e_a\to \int \rho(r)\,d\vecr =Ze$ is the total charge,
$\sum_a e_a \vecrho_a^2 \to \int r^2 \rho(r)\,d\vecr
 =e \langle Zr_E^2 \rangle$
is the electric mean square radius, and
$F(-\vecq^2) = Z - \frac16 \langle Zr_E^2 \rangle \vecq^2 + \cdots$
is the electric form factor.  Of course, the response of the extended
system to the external field is determined by the potential
and its derivatives together with the corresponding moments of the
charge distribution.

   A relativistic generalization to the case of an extended pion
suggests the following substitution for the vector potential in
Eq.\ (\ref{covder})
\beq
\label{subs}
    Z A_\mu(x) \to F(-\partial^2) A_\mu(x)
\eeq
which leads to the effective Lagrangian\footnote{Since we want to apply
the Lagrangian for the case of space-like virtual photons, we assume
the form factor to be real.}
\beq
\label{L-ff}
   \L_{\mbox{\footnotesize eff}}(\partial_\mu\phi,\phi)=
   [\partial_\mu + ie F(-\partial^2) A_\mu] \phi
   [\partial^\mu  -ie F(-\partial^2) A^\mu] \phi^\dagger
        - M^2 \phi\phi^\dagger,
\eeq
and to the first-order electromagnetic vertex
\beq
\label{emv}
  \Gamma_\mu(p',p) = (p'+p)_\mu F(q^2), \quad q=p'-p, \quad F(0)=Z.
\eeq
   Note that Eq.\ (\ref{L-ff}) is no longer invariant under the
canonical gauge transformation of Eq.\ (\ref{gauge}) and, correspondingly,
the electromagnetic vertex of Eq.\ (\ref{emv}) does not satisfy the
Ward-Fradkin-Takahashi identity \cite{Ward_50,Fradkin_56,Takahashi_57}
\beq
   q_\mu \Gamma^\mu(p',p)\neq Z[\Delta^{-1}(p')-\Delta^{-1}(p)],
\eeq
   where $\Delta(p)=1/(p^2-M^2)$ is the free propagator of
${\cal L}_0(\partial_\mu\phi,\phi)$.
   In fact, a different gauge transformation
\beq
\label{gauge-F}
   A_\mu(x) \mapsto A_\mu(x) + \partial_\mu \Lambda(x), \qquad
   \phi(x) \mapsto \exp[-ieF(-\partial^2)\Lambda(x)]\phi(x),
\eeq
   defines a local realization of the symmetry group U(1), i.e.,
under two successive transformations described by smoothly
varying functions $\Lambda_1$ and $\Lambda_2$, the fields transform as
\beqn
      A_\mu   &\mapsto&  A_\mu+\partial_\mu \Lambda_1
   \mapsto (A_\mu +\partial_\mu\Lambda_1)+\partial_\mu\Lambda_2
           =A_\mu+\partial_\mu(\Lambda_1+\Lambda_2),
\nn
     \phi   &\mapsto&
     \exp[-ieF(-\partial^2)\Lambda_1]\phi \mapsto
     \exp[-ieF(-\partial^2)\Lambda_2]\exp[-ieF(-\partial^2)\Lambda_1]\phi
\nn  &=&
    \exp[-ieF(-\partial^2)(\Lambda_1+\Lambda_2)]\phi.
\eeqn
   Accordingly, we define a {\em noncanonical} covariant derivative as
\beq
    \tilde{D}_\mu \phi=[\partial_\mu + ie F(-\partial^2) A_\mu] \phi,
\eeq
   such that $\tilde{D}_\mu \phi$ transforms in the same way
as $\phi$  under Eq.\ (\ref{gauge-F}) and the effective Lagrangian
\beq
\label{L-ff2}
   \L_{\mbox{\footnotesize eff}}(\partial_\mu\phi,\phi)
        =\L_0(\tilde{D}_\mu\phi,\phi)
        =\tilde{D}_\mu \phi (\tilde{D}^\mu \phi)^\dagger
           - M^2 \phi\phi^\dagger
\eeq
remains invariant under Eq.\ (\ref{gauge-F}).
   We stress that from a formal point of view {\em any} (real) function
represented by a power series would yield a realization of gauge
invariance and that the choice of the electromagnetic form factor
is entirely motivated on physical grounds through Eqs.\ (\ref{w2})
and (\ref{subs}).

    Although a description of a finite-size pion based on Eq.\ (\ref{L-ff})
is mathematically consistent, it turns out to be inconvenient as soon as
interactions of several particles with different form factors are considered.
   For instance, even a simple local interaction of
fields $\phi_a$ ($a=1,\cdots, n$) of the form
\beq
\label{L-local}
   \L_{\mbox{\footnotesize int}}(\phi_a) = g \prod_a \phi_a,
\eeq
   with a vanishing net charge associated with the product of fields,
is no longer automatically gauge invariant as soon as
different fields transform with different form
factors $F_a$.
   Under the gauge transformation of Eq.\ (\ref{gauge-F}), the interaction
Lagrangian, Eq.\ (\ref{L-local}),
picks up a space-time-dependent phase factor
\beq
  \prod_a \exp[-ieF_a(-\partial^2)\Lambda(x)] \ne 1.
\eeq
  On the other hand, this would not happen for the transformation law
of Eq.\ (\ref{gauge}), because
\beq
  \prod_a \exp[-ieZ_a\Lambda(x)] =1,
\eeq
provided the electric charge is conserved at the vertex, i.e.,
$\sum_a Z_a=0$.

   We therefore redefine the field as follows \cite{lvov87}:
\beq
   \phi(x) = \varphi(x) \exp[ief(-\partial^2)\partial_\mu A^\mu(x)],
\eeq
where\footnote{Recall our assumption that $F$ can be expanded in an
absolutely convergent series.}
\beq
     f(q^2) = \frac{1}{q^2}[F(q^2)-Z]
\eeq
   generates a new field which transforms
canonically under Eq.\ (\ref{gauge-F}), i.e.
\beqn
     \varphi  &=&  \exp[-ie f(-\partial^2)\partial_\mu A^\mu]\phi\\
          &\mapsto&
    \exp[-ief(-\partial^2)(\partial_\mu A^\mu+\partial^2\Lambda)]
    \exp[-ie F(-\partial^2)\Lambda]\phi
\nn   &=&
    \exp[-ief(-\partial^2)(\partial_\mu A^\mu+\partial^2\Lambda)]
    \exp\{ -ie [Z - \partial^2 f(-\partial^2)] \Lambda \}\phi
\nn   &=&
    \exp[-ief(-\partial^2)\partial_\mu A^\mu]\exp(-ieZ\Lambda)\phi
          =\exp(-ieZ\Lambda)\varphi.
\eeqn
   Rewritten in terms of $\varphi$, the Lagrangian (\ref{L-ff2}) reads
\beq
\label{L-ff'}
   \L_0(D^f_\mu\varphi,\varphi) = D^f_\mu \varphi (D^{f\mu} \varphi)^\dagger
    - M^2 \varphi\varphi^\dagger,
\eeq
where the second (noncanonical) covariant derivative
$D_\mu^f \varphi$ is defined as
\beq
     D_\mu^f\varphi = [\partial_\mu + ieZ A_\mu
    + ie f(-\partial^2)\partial^\nu \F_{\mu\nu}]\varphi,
\eeq
and $\F_{\mu\nu} = \partial_\mu A_\nu - \partial_\nu A_\mu$.
   This second definition ensures canonical gauge invariance and
simultaneously accounts for the finite size of the particle.

   Organized in powers of the elementary charge, the Lagrangian  can
be expressed in terms of the canonical covariant derivative as
\beq
   \L_0(D_\mu^f\varphi,\varphi) = \L_0(D_\mu \varphi,\varphi)
 + \L_1(D_\mu \varphi,\varphi) + \L_2(\varphi).
\eeq
$\L_0$ and
\beq
   \L_1(D_\mu \varphi,\varphi) = -ie [\varphi^\dagger \tensor D_\mu \varphi] \,
        f(-\partial^2)\partial_\nu \F^{\mu\nu}
\eeq
generate the electromagnetic vertex
\beq
  \Gamma_\mu(p',p) = (p'+p)_\mu F(q^2)
   - q_\mu (p^{\prime 2} - p^2) f(q^2), \quad q=p'-p,
\eeq
which now satisfies the Ward-Fradkin-Takahashi identity
\beq
  q_\mu \Gamma^\mu(p',p) = Z (p^{\prime 2} - p^2) =
    Z [\Delta^{-1}(p') - \Delta^{-1}(p)]
\eeq
with the free propagator.
The last term
\beq
   \L_2(\varphi) = e^2 [f(-\partial^2)\partial_\nu \F^{\mu\nu}]
       [f(-\partial^2)\partial^\rho \F_{\mu\rho}] \varphi\varphi^\dagger
\eeq
vanishes, when at least one of the photons is real, thus
satisfying $\partial_\nu \F^{\mu\nu} = 0$.

In the following we {\em define} the (generalized) Born terms
of the virtual Compton scattering amplitude
as the VCS amplitude constructed from either of the Lagrangians of Eqs.\
(\ref{L-ff}) and (\ref{L-ff'}).
   According to the equivalence theorem of Lagrangian field theory
\cite{Chisholm_61,Kamefuchi_61}, the scattering amplitude does not
depend on the Lagrangian used as long as all external pions are on shell.
   To be specific, one obtains
\beq
   T_{\mu\nu}^{\rm Born} = e^2 F(q^2) F(q^{\prime 2})
   \Bigg[ 2g_{\mu\nu}
   - \frac{(2p+q )_\mu (2p'+q')_\nu} {(p+q )^2 - M^2}
   - \frac{(2p-q')_\nu (2p'-q )_\mu} {(p-q')^2 - M^2} \Bigg].
\label{Born}
\eeq
In fact, such a form of the Born amplitude is standard for a
discussion of structure-dependent characteristics of the target,
and we have shown that this is a very natural
generalization of the Born amplitude for a point-like particle
to the case of a finite-size particle.
   As discussed in Ref.\ \cite{fearing98} in detail, such a
generalization incorporates all low-energy singularities of the total
VCS amplitude so that the non-Born part of the
amplitude can be expanded in powers of small photon momenta giving rise
to (generalized) polarizabilities.

\section{Low-momentum expansion and generalized dipole polarizabilities}
\label{sec:LEX}

   A well-known, general method of obtaining the low-energy expansion of a
reaction amplitude consists of expanding invariant amplitudes free from
kinematical singularities and constraints in a power series
(see e.g.\ \cite{bardeen68,choudhury68}).
Following this method, we first decompose the invariant amplitudes
$B_i(\nu^2, q\cdot q', q^2 + q^{\prime 2}, q^2 q^{\prime 2})$
into generalized Born and non-Born contributions,
\beq
    B_i = B_i^{\rm Born} + B_i^{\rm NB}, \quad i=1,\cdots,5.
\eeq
The generalized Born, or convection contribution is associated with a
set of diagrams describing single-pion exchanges in $s$-
and $u$-channels with $\gamma\pi\pi$ vertices taken in the on-shell
regime.
   As we have seen in the previous section, additional non-pole terms
are necessary to render the generalized Born terms gauge invariant.
   The thus constructed amplitude possesses all the symmetries of the
total amplitude $T_{\rm VCS}$ and contains all singularities of $T_{\rm VCS}
$ at low energies.  Using the specific form of the Born amplitude
$T_{\rm VCS}^{\rm Born}$ given in the previous section, we find the
(generalized) Born parts of the invariant amplitudes $B_i$,
\beq
\label{B-born}
   B_1^{\rm Born} = (q\cdot q') C, \quad
   B_2^{\rm Born} = -4C, \quad
   C = \frac{2e^2 F(q^2) F(q^{\prime 2})} {(s-M^2)(u-M^2)},
\eeq
and $B_i^{\rm Born}=0$ for $i=3,4,5$.

   At energies below inelastic thresholds, the non-Born parts of $B_i$ are
regular functions of the kinematical variables.
   They determine the deviation of $T_{\rm VCS}$ from its Born value of
Eq.\ (\ref{Born}).
   In particular, when the momenta of both photons are small,
$q \sim q' \to 0$, one obtains
\beq
\label{LET-2}
     T_{\rm VCS} = T_{\rm VCS}^{\rm Born}
    + \frac12 F^{\mu\nu} F_{\mu\nu}' b_1(0)
      + (P_\mu F^{\mu\nu})(P^\rho F_{\rho\nu}') b_2(0) + \O(q^4),
\eeq
where the constants $b_{i}(0) \equiv B_{i}^{\rm NB}(0,0,0,0)$, $i=1,2$,
can be related to the ordinary electric and magnetic dipole
polarizabilities of low-energy real Compton scattering,
\beq
\label{real-a,b}
     8\pi M \bar\alpha = -b_1(0) - M^2 b_2(0), \quad
     8\pi M \bar \beta =  b_1(0).
\eeq
Equations (\ref{LET-2}) and (\ref{real-a,b}) provide a Lorentz-invariant
form of the low-energy theorem for virtual Compton scattering up to
second order in the photon momenta (for the case of the nucleon, see
Ref.\ \cite{Scherer_96}).

Now, following the original idea of Guichon {\it et al.} \cite{guichon95},
we consider the case when the final photon is real and has a very small
momentum $q'\to 0$, whereas the initial photon momentum $q$ is allowed
to be virtual and is not necessarily small.
   As may be seen from Eq.~(\ref{ampl}), the amplitude of
$\gamma^\ast \pi \to \gamma \pi$ with a real final photon is determined
by three invariant amplitudes $B_1$, $B_2$, and $B_3$,\footnote{The
definition of $P$ of Eq.\ (\ref{pqr}) differs by a factor of $1/2$ from
Ref.\ \cite{unkmeir00}.}
\beq
\label{T-real}
     T_{\rm VCS} = \frac12 F^{\mu\nu} F_{\mu\nu}' B_1
      + (P_\mu F^{\mu\nu})(P^\rho F_{\rho\nu}') B_2
      + (P^\nu q^\mu F_{\mu\nu}) (P^\sigma q^\rho F_{\rho\sigma}') B_3.
\eeq
This equation has a particularly simple form in the pion Breit frame
(PBF) defined by $\vecP=0$, i.e., $\vecp=-\vecp'$,
in which the initial and final
pion are treated on a symmetrical footing.
   Introducing the Fourier components of the electric and magnetic
fields as
\beqn
\label{fields}
   & \vecE = i(q_0\vecepsilon - \vecq\epsilon_0),
\quad
    \vecB = i\vecq\times\vecepsilon,
\nn
   & \vecE' = -i(q'_0\vecepsilon^{\,\prime *}
        - \vecq^{\,\prime}\epsilon_0^{\prime *}),
\quad
    \vecB' = -i\vecq^{\,\prime}\times\vecepsilon^{\,\prime *},
\eeqn
we can rewrite Eq.~(\ref{T-real}) in the Breit frame as
\beqn
\label{T-Breit}
  T_{\rm VCS} &=& \left[(\vecB\cdot \vecB') B_1
      - (\vecE\cdot \vecE') (B_1 + P^2 B_2)
      + (\vecE\cdot \vecq) (\vecE'\cdot \vecq) P^2 B_3\right]_{PBF}.
\eeqn
  In general, even after subtraction of the singular (Born) parts of the
amplitudes, all $B_i^{\rm NB}$ still depend on the
photon energies and the scattering angle and thus describe a series of
multipoles and dispersion effects.
   However, in the limit of $q'\to 0$, only scalar structure functions
depending on $q^2$ survive:
\beq
  b_i(q^2) = B_i^{\rm NB}(0,0,q^2,0).
\eeq
These functions yield the non-Born parts of the coefficients in
(\ref{T-Breit}) which we interpret as generalized  electric and
magnetic dipole polarizabilities:
\beqn
\label{polars}
    8\pi M \beta(q^2) &=&  b_1(q^2), \nn
    8\pi M \alpha_T(q^2) &=& -b_1(q^2) - \left(M^2-\frac{q^2}{4}
    \right) b_2(q^2), \nn
    8\pi M \alpha_L(q^2) &=& -b_1(q^2) - \left(M^2 -\frac{q^2}{4}\right)
    [b_2(q^2) + q^2 b_3(q^2)],
\eeqn
where $P^2$ has been taken in the limit $q'=0$ as well,
i.e.,  $P^2= M^2 - q^2/4$.
   If the initial virtual photon has a
transverse polarization in the Breit frame ($\vecE = \vecE_T \perp
\vecq$), the pion response to the transverse electric and magnetic
fields is determined by the (generalized) transverse electric and
magnetic polarizabilities:
\beq
\label{T-T}
  (T_{\rm VCS}^{\rm NB})_T = 8\pi M [
      (\vecE_T\cdot \vecE') \alpha_T(q^2)
    + (\vecB\cdot \vecB') \beta(q^2)] +
    \mbox{(higher orders in $q'$)}.
\eeq
For a longitudinal polarization ($\vecE = \vecE_L \parallel\vecq$)
in the Breit frame,
\beq
\label{T-L}
  (T_{\rm VCS}^{\rm NB})_L = 8\pi M
     (\vecE_L\cdot \vecE') \alpha_L(q^2) + \O(q^{\prime 2}).
\eeq
In the real-photon limit, $q^2 \to 0$, we have
\beq
\label{rpl}
      \beta(0) = \bar\beta, \quad
      \alpha_L(0) = \alpha_T(0) = \bar\alpha.
\eeq
All thus defined polarizabilities are functions of $q^2$ free from
kinematical singularities. In particular, for pion, kaon, or
nucleon targets they are regular functions below the two-pion threshold,
$q^2 < 4m_\pi^2$, and this region includes all space-like momenta.

   We will now interpret $\alpha_L(q^2)$, $\alpha_T(q^2)$, and
$\beta(q^2)$ by means of a semi-classical qualitative picture.
   To that end, let us consider a system of $n$ polarizable (neutral)
constituents at positions $\vecr_a$ ($a=1,\cdots, n$) with electric
polarizabilities $\alpha_{ij}^a$.
   We allow the constituents to be anisotropic, i.e., the
polarizability tensors are symmetric,  $\alpha_{ij}^a=\alpha_{ji}^a$, but
not necessarily diagonal ($\propto \delta_{ij}$).
   The system will respond to an external electric field $\vecE(t,\vecr)$
by acquiring a dipole moment%
\footnote{
The factor of $4\pi$ is related with the (standard)
use of Gaussian units for the polarizabilities
but natural units for charges and fields.}
\beq
   D_i(t) = 4\pi  \sum_{a=1}^n  \alpha_{ij}^a\, E_j(t,\vecr_a).
\eeq
   Slow oscillations of this dipole moment generate radiation of an outgoing
long-wavelength electromagnetic wave $\vecE'(t,\vecr)$
through the interaction $-\vecD\cdot\vecE$.
   For an incoming plane wave with momentum $\vecq$, i.e.,
$\vecE(\vecr) = \vecE \exp(i\vecq\cdot\vecr)$,
the amplitude for a transition to an outgoing plane wave with a very small
momentum (viz.\ $q' \rho \ll 1$,  where $\rho$ characterizes the
 system's extension) reads
\beq
    f_{fi} = 4\pi  \sum_{a=1}^n \alpha_{ij}^a\,
   \exp(i\vecq\cdot\vecrho_a) \,E_i E_j' \, ,
\eeq
where $\vecrho_a = \vecr_a - \vecR$ are the positions of the constituents
with respect to the center of mass $\vecR$ of the system.
   The continuum limit of a system with spherical symmetry
must be of the form
\beq
\label{polarizability-tensor}
     \sum_{a=1}^n \alpha_{ij}^a\, \exp(i\vecq\cdot\vecrho_a)
           \to \alpha_{ij}(\vecq) =
     \alpha_L(q) \hat q_i \hat q_j
      + \alpha_T(q) (\delta_{ij} - \hat q_i \hat q_j),
\eeq
where $\alpha_L$ and $\alpha_T$ do not depend on the direction $\hat q$
of $\vecq$. In this way we recover the structure of the VCS
amplitude given by Eqs.\ (\ref{T-T}) and (\ref{T-L}).

   If the system under consideration is exposed to a static and uniform
external electric field $\vecE'$, an electric polarization
$\vecPol$ is generated which is related to the {\em density} of the induced
electric dipole moments:
\beq
\label{d-induced}
     \Pol_i(\vecr) = 4\pi\sum_{a=1}^n\alpha_{ij}^a \,
      \delta^3(\vecr-\vecr_a) \,E_j'
        \equiv 4\pi\alpha_{ij}(\vecr - \vecR) \,E_j'  \, .
\eeq
   The tensor $\alpha_{ij}(\vecr)$ is nothing else
but the Fourier transform of
the polarizability tensor of Eq.\ (\ref{polarizability-tensor}):\footnote{
We do not use different symbols for a function $f(t)$ and its Fourier
transform $\tilde{f}(\omega)$.}
\beq
\label{alpha_ijr}
     \alpha_{ij}(\vecr) = \int \alpha_{ij}(\vecq)
     \exp(-i\vecq\cdot\vecr) \,\frac{d\vecq}{(2\pi)^3}.
\eeq
   If we define
\beq
    B(q)=\frac{\alpha_L(q)-\alpha_T(q)}{q^2},
\label{bq}
\eeq
  Eq.\ (\ref{polarizability-tensor}) can be rewritten as
\beq
   \alpha_{ij}(\vecq) = \alpha_T(q) \delta_{ij} + B(q) q_i q_j
\eeq
such that the Fourier transformation results in
\beq
   \alpha_{ij}(\vecr) = \alpha_T(r) \delta_{ij} - \nabla_i\nabla_j B(r).
\label{aaa}
\eeq
   Because of $B=B(r)$, the second term of Eq.\ (\ref{aaa}) reads
\beq
   - \nabla_i\nabla_j B(r) =
-\left(\delta_{ij} - \frac{r_i r_j}{r^2}\right)\frac{B'(r)}{r}
        - \frac{r_i r_j}{r^2} B''(r).
\label{ninjb}
\eeq
   On the other hand, from Eq.\ (\ref{bq}) written in the form
$q^2 B(q) = \alpha_L(q)-\alpha_T(q)$, one obtains for the
Fourier transform
\beq
     -\nabla^2 B(r) = -B''(r) - \frac{2}{r}B'(r) = \alpha_L(r)-\alpha_T(r).
\label{bbb}
\eeq
   Eliminating $B''(r)$ from Eq.\ (\ref{ninjb}) allows one to
rewrite Eq. (\ref{aaa}) as
\beq
   \alpha_{ij}(\vecr) = \alpha_L(r) \frac{r_i r_j}{r^2}
      + \alpha_T(r) \Big(\delta_{ij} - \frac{r_i r_j}{r^2}\Big)
      + \frac{3r_i r_j - r^2 \delta_{ij}}{r^3} B'(r).
\label{ccc}
\eeq
   Finally, the last term of Eq.\ (\ref{ccc}) is determined by
reexpressing Eq.\ (\ref{bbb}) as
\beq
    \frac{d}{dr}[r^2 B'(r)] = - r^2 [\alpha_L(r)-\alpha_T(r)],
\eeq
   which, assuming the boundary condition
$\lim_{r\to\infty} r^2 B'(r)=0$, is solved by%
\footnote{Instead of Eq.\ (\ref{ddd}) we could also use
$$
    r^2 B'(r) = - \int_0^r r'^2 [\alpha_L(r')-\alpha_T(r')]\,dr',
$$
resulting from the boundary condition $\lim_{r\to 0} r^2 B'(r)=0$.
   Both results are identical, because
$$
    \int_0^\infty 4\pi r^2 [\alpha_L(r)-\alpha_T(r)]\,dr \equiv
    \alpha_L(q=0) - \alpha_T(q=0) = 0,
$$
where we made use of Eq.\ (\ref{rpl}).}

\beq
    r^2 B'(r) = \int_r^\infty r'^2 [\alpha_L(r')-\alpha_T(r')]\,dr'.
\label{ddd}
\eeq
   In other words, given the Fourier transforms
\beq
\label{alphaltr}
     \alpha_{L,T}(r) \equiv \int
     \alpha_{L,T}(q) \exp(-i\vecq\cdot\vecr) \,\frac{d\vecq}{(2\pi)^3},
\eeq
 the density of the full electric polarizability of the system,
$\alpha_{ij} = \sum_a \alpha_{ij}^a$, can be reconstructed as
\beqn
\label{alphaijr}
     \alpha_{ij}(\vecr) &=&
     \alpha_L(r) \hat r_i \hat r_j
      + \alpha_T(r) (\delta_{ij} - \hat r_i \hat r_j)
      + \frac{3\hat r_i \hat r_j - \delta_{ij}}{r^3}
        \int_r^\infty [\alpha_L(r')-\alpha_T(r')]\,r'^2\,dr'.
\eeqn

   In this context, it is important to realize that both longitudinal and
transverse polarizabilities, $\alpha_L$ and $\alpha_T$, respectively, are
needed to fully recover the electric polarization $\vecPol$.
   The longitudinal polarizability is special, though, because it
completely specifies the induced polarization charge density of the system,
\beq
    \delta\rho(\vecr) = -\vecnabla\cdot\vecPol(\vecr)
    = -4\pi (\vecE'\cdot\vecnabla) \alpha_L(r),
\eeq
where we made use of
\beq
   \nabla_i\alpha_{ij}(\vecr) = \nabla_j\alpha_L(r)
\eeq
which follows from Eqs.\ (\ref{polarizability-tensor}) and
(\ref{alpha_ijr}).
   Combining partial integration with the divergence theorem, one finds
for the Fourier transform of the induced polarization charge
\beq
\label{P}
    \delta\rho(\vecq) \equiv
   \int \delta\rho(\vecr) \exp(i\vecq\cdot\vecr) \, d\vecr
    = i\vecq\cdot\vecPol(\vecq)
       = 4\pi i\alpha_L(q) \, \vecq\cdot\vecE'.
\eeq
   Such an induced charge density is the source of a longitudinal
electric (Coulomb) field and thus generates an effective coupling of the type
$\vecE_L\cdot \vecE'$.

   At the same time, the transverse polarizability $\alpha_T$ describes
rotational displacements of charges which do not contribute to
$\delta\rho(\vecr)$.
   They can generate electric fields only for a finite frequency,
$q'_0 \ne 0$.
   Therefore, in the limit $q_0' \to 0$ the effective coupling
$\vecE_T\cdot \vecE'$ should vanish faster than $\vecE_L\cdot \vecE'$.
   This is indeed the case, as we will see below.

The relation (\ref{P}) suggests an intimate connection between
the longitudinal polarizability $\alpha_L$ and the charge-density
operator of the system.
   In a forthcoming publication we will verify this in a quantum-mechanical
and fully relativistic framework.

   Similar considerations apply to the magnetic part of the VCS amplitude.
   In this case the electric polarizabilities $\alpha_{ij}^a$
should be replaced by the magnetic ones, $\beta_{ij}^a$.
   Since the magnetic induction is always transverse
(i.e., $\vecB\cdot\vecq=0$), terms containing $\hat q_i \hat q_j$ in the
magnetic analogue of Eq.\ (\ref{polarizability-tensor}) do not enter any
observable and can thus be omitted.
   Hence, the unobservable ``longitudinal" magnetic
polarizability $\beta_L(q)$ can for all $q$ be chosen to be identical with the
transverse one, $\beta_T(q)$ rather than only at the point $q=0$, where
the equality $\beta_L(0)=\beta_T(0)$ is dictated by analyticity.
   With this choice $\beta_{ij}(\vecq) = \beta(q)\,\delta_{ij}$, and
the analogue of Eq.\ (\ref{alphaijr}) reads
\beq
\label{betaijr}
    \beta_{ij}(\vecr)=\beta(r)\delta_{ij}.
\eeq
   Then the magnetization $\vecMag$ induced by the uniform external magnetic
field and the corresponding induced electric current
$\delta\vecj(\vecr) = \vecnabla\times\vecMag(\vecr)$
are, up to an arbitrary gradient, given in terms of the density of the
magnetic polarizability as
\beq
    \vecMag(\vecr) = 4\pi\beta(\vecr - \vecR) \,\vecB',
\eeq
where the Fourier transform of $\beta(r)$ is nothing but the generalized
magnetic polarizability $\beta(q)$:
\beq
     \beta(r) = \int \beta(q)
    \exp(-i\vecq\cdot\vecr) \, \frac{d\vecq}{(2\pi)^3}.
\eeq

   Let us conclude this section by recalling that Eq.\ (\ref{T-T}) was
obtained by keeping the lowest multipolarities of the
final (soft) photon.
   This, however, leads to different powers of the
photon momentum $q'$ in such an expansion.
   For $q' \to 0$, the energy of the initial photon in the Breit frame
vanishes as well, because $q_0=q'_0$ in that frame.
   It then follows from Eq.\ (\ref{fields}) that the transverse
electric field $\vecE_T$ is of higher order in $q'$ than the
(transverse) magnetic field $\vecB$.
   This is also clear from the relation
$\vecq^2 \vecE_T = -q_0 \vecq\times\vecB$.
   When only terms up to order $\O(q')$ are retained, the transverse
electric field does not contribute!
   In order to translate Eq.\ (\ref{T-Breit}) into a power expansion of the
non-Born part of the VCS amplitude up to $\O(q^{\prime 2})$,
one has to add two more terms proportional to
$(q\cdot q') [\partial B_1^{\rm NB}/ \partial(q\cdot q')]_{q'=0}$ and
to a similar derivative of the function $B_1 + P^2 B_2$.
   These terms introduce an additional angular dependence and therefore
higher multipoles (quadrupoles).

    The physical process of photon electroproduction
$e(k) + h \to e'(k') + h' + \gamma(q')$ consists of the Bethe-Heitler
contribution, in which the real final photon is emitted by the initial
and final electrons, respectively, and the VCS contribution.
   The virtual photon of the VCS part, $\gamma(q)$,
is described in terms of the polarization vector
\beq
\label{epsilon-electron}
    \epsilon_\mu = \frac{e}{q^2} \bar u_e(k')\gamma_\mu u_e(k),
   \quad q = k - k',
\eeq
which is determined by the electron-scattering kinematics.
   For $q'\to 0$, such an $\epsilon_\mu$ remains finite.
   Therefore, the transverse electric field
$\vecE_T = iq_0\vecepsilon_T$ created by the electron transition
current is of order $\O(q')$ in the Breit frame and is suppressed
in comparison with the magnetic and longitudinal electric fields
generated by the current.
   As an immediate consequence the non-Born part
of the VCS amplitude to order $\O(q')$ is characterized by {\em two}
structure functions [viz.\ $\alpha_L(q^2)$ and $\beta(q^2)$] rather
than by all three functions appearing in the dipole approximation.

   Although we arrived at this conclusion by considering the VCS amplitude in
the Breit frame [$\vecp=(\vecq'-\vecq)/2$], it is clear that
two independent structure functions
characterize the amplitude to order $\O(q')$ in any other frame such
as, for example, the center-of-mass (c.m.) frame.
   This is true because the amplitude itself is Lorentz invariant and
at the same time the real-photon energy, $\omega'=|\vec{q}\,'|$,
remains of the same order $\O(q')$ for any {\em finite} Lorentz boost.

   The above consideration gives a transparent explanation of the theorem
established in Ref.\ \cite{drechsel97}, which states that there are
only two independent structure functions which determine $\O(q')$ terms
in the so-called full amplitude $T_{\rm FVCS}^{\rm NB}$ of virtual
Compton scattering for a spin-0 particle in the c.m.\ frame.
   Here, ``full'' refers to the fact that the polarization and the
intensity of the initial photon are given by
Eq.\ (\ref{epsilon-electron}).

   In terms of the notation introduced by Guichon {\it et al.}
\cite{guichon95}, this theorem establishes a linear relation between the
c.m.\ generalized polarizabilities $P^{(01,01)0}$, $P^{(11,11)0}$, and
$\hat P^{(01,1)0}$, leaving only two of them independent.
   In the c.m.\ frame, the transverse electric field is not vanishing, and the
above theorem can be re-phrased as establishing a linear combination of the
$\vecE_L$, $\vecE_T$ and $\vecB$ responses in the c.m.\ frame which
vanishes when $q'=0$.  This is just the $\vecE_T$ response in the
Breit frame.

   The explicit relations between the c.m.\ polarizabilities
and the quantities $\alpha_L(q^2)$ and $\beta(q^2)$ appearing
at order $\O(q')$ read
\beqn
\label{ab-vs-P}
       \alpha_L(q^2) &=& -\frac{e^2}{4\pi}~
   \sqrt{\frac{3E_{\rm cm}}{(2J+1)M}}~ P^{(01,01)0}(q_{\rm cm}),
\nn
       \beta(q^2) &=& -\frac{e^2}{8\pi}~
   \sqrt{\frac{3E_{\rm cm}}{(2J+1)M}}~ P^{(11,11)0}(q_{\rm cm}),
\eeqn
where $E_{\rm cm}=M-q^2/(2M)$ and $q_{\rm cm}=\sqrt{-q^2+ q^4/(4M^2)}$
denote the energy and the absolute value of the three-momentum of the
initial pion in the c.m.\ frame at threshold ($q'=0$).
   The spin factor $2J+1$ removes a related spin
dependence hidden in the quantities $P^{(\rho' L',\rho L)S}$
and is needed when our ``pion" represents a spin-averaged hadron of
spin $J\ne 0$.

   When considering the Fourier transforms,
the additional factor of $\sqrt E_{\rm cm}$ in Eq.\ (\ref{ab-vs-P}) and the
use of the Breit-frame momentum transfer $q_{\rm Breit}=\sqrt{-q^2}$
instead of the c.m. momentum transfer $q_{\rm cm}$ will generate a difference
for the spatial distributions, especially for such a light particle as the
pion.
   From the analogy with the well-known case of electromagnetic form factors,
where spatial distributions are obtained using the Breit-frame variables,
we expect that a meaningful Fourier transformation in the case of
the generalized polarizabilities also requires the Breit frame.
   Indeed, the only difference between the kinematics of the reaction of VCS,
$\gamma^*\pi\to\gamma'\pi$, and the kinematics of the reaction
$\gamma^*\pi\to\pi$, in which the form factors are studied,
originates in the presence of an additional photon $\gamma'$ which carries a
vanishing momentum $q'=0$.

   In analogy to Eq.\ (\ref{ampl}), the structure-dependent effects as seen
in VCS with one (soft or hard) space-like virtual and one soft real photon
can be encoded in the following effective Lagrangian:
\beq
  \L_{\rm polariz} = \frac14 \b_1 \F^{\mu\nu} \F_{\mu\nu} \phi^\dagger\phi
    + \frac12 [ \b_2 \F^{\alpha\mu} \F_{\beta\mu}
    -  2\b_3 \F^{\alpha\mu}\, \partial_\mu\partial^\nu \F_{\beta\nu} ]
       \P_\alpha \P^\beta \phi^\dagger \phi,
\eeq
where $\b_i$ are differential operators acting on
the electromagnetic fields which are determined by their Fourier components
$b_i(q^2)$ [see Eq.\ (\ref{polars})].
   The above Lagrangian contains all possible gauge-invariant terms
involving at least one field strength tensor $\F_{\mu\nu}$ without
derivatives [$\O (\omega')$!].

\section{Spatial distributions in chiral perturbation theory}
\label{sec:spatial}

\subsection{Preliminary remarks}

   In accordance with the ideas presented in the previous section, we now
consider the Fourier transforms of the $q$-dependent polarizabilities,
generically denoted by $F(q^2)$,
and discuss the corresponding spatial distributions $F(r)$.

   There is a well-known objection against a straightforward
interpretation of such $F(r)$ as a spatial distribution.
   The argument is related to the fact that the velocities of the target
before and after the interaction with the virtual photon depend on the
photon momentum.
   If we think of the target as a composite system of, say,  quarks we would
expect that the matrix element
\beq
\label{matrix-element}
    \int \psi^\dagger_f(\vecR',\tau')
    \langle \vecR',\tau'|\O | \vecR,\tau \rangle \psi_i(\vecR,\tau)
    \,d\vecR \,d\tau \,d\vecR' \,d\tau'
\eeq
of a transition operator like $\O=\int j^\mu(x)A_\mu(x)\,d^3x$
for a one-photon reaction%
\footnote{Another example would be
$\O=\int\!\!\int j^\nu(x')A_\nu'(x')
   G_E(x',x) j^\mu(x)A_\mu(x)\,d^3x\,d^3x'$
involving two currents, where $G_E$ is the Green function.}
depends on both the internal (Lorentz-invariant) variables
$\tau$ of the pion {\em and} on the pion's center-of-mass variable $\vecR$
(cf.\ Ref.\ \cite{krajcik74}).
   Since a relativistic wave function $\psi(\vecR,\tau)$, in general,
does not factorize into a product of the type
$\phi(\tau)\exp(i\vecp\cdot\vecR)$,
where $\phi(\tau)$ denotes a $\vecp$-independent internal wave function,
some part of the full $q$-dependence of
the transition matrix element may be related with the c.m., partly
kinematical effects of $\vecp$ on $\phi(\tau)$.
   We come closest to associating the Fourier transform $F(r)$
of the {\em full} matrix element (\ref{matrix-element}) with the {\em
internal} spatial structure of the particle by evaluating this matrix
element in the Breit frame,
in which the pion is at rest on the average, i.e.,
$\vecp + \vecp'=0$.

   There is a simple phenomenological argument
suggesting that the generalized magnetic polarizability $\beta(q^2)$
defined according to Eq.~(\ref{polars}) is indeed only related to
the internal structure.
   The point is that this quantity is a function
of $q^2$ without kinematical singularities.
   In other words, any irregularity in its $q^2$ behavior is not caused by a
Lorentz contraction
and has nothing to do with singular $\gamma$ factors or with quantities like
$P_0=\frac12\sqrt{4M^2+\vecq^2}$.
    Moreover, even singularity-free
quantities like $P_0^2=P^2$ should be irrelevant, because the momentum
scale on which the amplitude $b_1(q^2)$ changes has nothing to do with
the particle mass $M$ itself and is fully determined by interactions.

   For the other polarizabilities the situation may be more complicated.
   The sums $\alpha_L(q^2) + \beta(q^2)$ and $\alpha_T(q^2) + \beta(q^2)$
contain an overall factor of $P^2$ [see Eqs.~(\ref{polars})] which,
for the pion, introduces a small kinematical mass scale into
the $q^2$ behavior of these sums.
   So, perhaps a more meaningful consideration of spatial distributions
relating to $\alpha_L(r) + \beta(r)$ and $\alpha_T(r) + \beta(r)$ is obtained
with the factor $P^2/M^2$ excluded from $\alpha_L(q^2) + \beta(q^2)$ and
$\alpha_T(q^2) + \beta(q^2)$ before performing the Fourier transformations.
   In the following discussion we will not encounter this problem,
because in the theory considered, namely, ChPT at lowest nontrivial order,
either the sums $\alpha_L(q^2) + \beta(q^2)$ and
$\alpha_T(q^2) + \beta(q^2)$ are exactly zero
[for pions and kaons at $\O(p^4)$] or the particle mass $M$ is
considered to be infinite [for baryons in HBChPT at $\O(p^3)$].

   We will illustrate, by means of the more familiar example of form factors,
that associating a generic $F(r)$ with the internal structure of the
particle leads to a self-consistent picture and does not create any visible
problems even in the case of such a light particle as the pion, for
which the relativistic interlace of c.m.\  and internal variables is
maximal.
   To be specific, we will consider form factors calculated in the
framework of ChPT, mainly for two reasons.
   First, we want to discuss the generalized polarizabilities $F(q^2)$
predicted by the {\em same} theory in order to check that our consideration of
polarizabilities at scales $r\sim 1/m_\pi$ is not in conflict with
other observables.
   Second, at present ChPT is the best tool for describing hadron structure at
large scales and it is the only theory which agrees with the recent MAMI
data on generalized polarizabilities of the proton \cite{roche00}.

   We would like to mention the following technical aspect concerning
the Fourier transformation of a $q$ distribution obtained within ChPT.
   Such distributions are only reliably known for small momenta,
$q=\O(m_\pi)$.
   Moreover, a straightforward integration over $q$ does not exist,
because the integrand typically diverges for large values of $q$.
   We therefore enforce convergence by introducing a cutoff $\Lambda$.
   Clearly, such a cutoff disturbs the corresponding $r$ distributions
at distances $r\lesssim 1/\Lambda$ which are beyond the scope of
ChPT.
   On the other hand, one might expect that the results are cutoff independent
when $r \gg 1/ \Lambda$.
   In practice, we make use of a Gaussian cutoff and,
for any $q$-dependent form factor or polarizability $F(q^2)$, we
calculate $F(r)$ as
\beq
\label{Fourier-reg}
     F(r) = \lim_{\Lambda \to \infty} \,
     4\pi    \int_0^\infty F(-Q^2) \,\frac{\sin(Qr)}{Qr}\,
     \exp\Big({-}\frac{Q^2}{\Lambda^2}\Big)\,
     \frac{Q^2\,dQ}{(2\pi)^3}.
\eeq
   Depending on how small $r$ is, we have to choose $\Lambda$ large enough
in order to approach the limit of $\Lambda=\infty$.
   In particular, for all generalized polarizabilities considered
below we have found the regularized Fourier integral of
Eq.\ (\ref{Fourier-reg}) to be independent of $\Lambda$ for
$\Lambda \ge 30$ GeV within an accuracy better than 2\% even at
distances as short as $r=0.1$ fm.
   Stated differently, the cutoff $\Lambda=30$ GeV is sufficient to resolve
spatial distributions of polarizations at scales
$r \sim 0.1$ fm.
   In the case of electromagnetic form factors, having steeper spatial
distributions (see below), the 30 GeV cutoff is sufficient
for a good resolution up to distances of $r \sim 0.2$ fm.

   There is yet another way of calculating the Fourier integral of
Eq.\ (\ref{Fourier-reg}) based on a contour deformation in the
complex $Q$ plane.
   This method is applicable when the analytical continuation of $F(q^2)$
to time-like momenta is known as in the case of the ChPT predictions.
   Since all singularities
of $F(t)$ are located at real positive $t$, it is possible
to write a dispersion relation for $F$,\footnote{
Additional subtractions may be required resulting in additional polynomial
contributions.}
\beq
     F(q^2) = \frac{1}{\pi} \int_{t_{\rm min}}^\infty
                {\rm Im}\, F(t)\,\frac{dt}{t-q^2-i0^+}
\eeq
 which allows one to recast the Fourier integral for $F(r)$ at $r>0$ as a
superposition of Yukawa functions:
\beq
\label{Yukawa}
     F(r) = \frac{1}{4\pi^2 r}
          \int_{t_{\rm min}}^\infty e^{-r\sqrt t} \, \Im F(t) \,dt .
\eeq
   We made use of both methods and arrived at identical results for $F(r)$.

   It is worthwhile recalling that polynomial pieces in $F(q^2)$ generate
$\delta(r)$ terms or derivatives thereof in the Fourier transform $F(r)$.
   Such terms typically originate from higher-order counter terms in the
Lagrangian which are multiplied by a priori unknown low-energy
constants.
   However, in the Fourier transform, they do not contribute to
$F(r)$ at finite $r\ne 0$.
   In other words, the Fourier integral acts as a filter which only transmits
genuine effects of pion (or kaon) loops through their contributions to a
nonpolynomial part of $F(q^2)$ and to ${\rm Im}\, F(q^2)$, respectively.

\subsection{Form factors}

   As a first illustration, we briefly discuss the scalar and vector form
factors of the pion as obtained in two-flavor ChPT in the limit of
isospin symmetry.
   These form factors parametrize matrix elements of the scalar
density $S(x) \equiv \hat m[\bar{u}(x)u(x) + \bar{d}(x)d(x)]$ with
$\hat m = m_u=m_d$ and the isovector electromagnetic current
$j_\mu^V(x) \equiv \frac12 \bar q(x) \tau_3 \gamma_\mu q(x)$, respectively:
\beqn
    \langle \pi(p') | S(0) | \pi(p) \rangle &=& F_S(q^2),
\nn
    \langle \pi^+(p') | j_\mu^V(0) | \pi^+(p) \rangle
      &=& (p+p')_\mu F_V(q^2), \qquad q=p'-p.
\eeqn
Recently, the one-loop calculations of $F_S(q^2)$ and $F_V(q^2)$ by Gasser
and Leutwyler \cite{gasser84} have been extended to the two-loop level
${\cal O}(p^6)$  \cite{colangelo96,bijnens98}.

   In order to simplify the discussion, we perform two subtractions
in the form factors $F(t)$ and plot subtracted (and normalized) functions,
\beq
   \bar F(t) = \frac{1}{F(0)}
   \Big[F(t) - t F'(0) - \frac12 t^2 F''(0)\Big].
\eeq
   By that means we avoid polynomial contributions of ${\cal O}(p^4)$
and ${\cal O}(p^6)$, which depend on low-energy constants,
and emphasize the pieces originating from pion loops.%
\footnote{Alternatively, we could keep the polynomial contribution
of the pion loops. However, in that case the result would still depend
on the renormalization condition chosen.}
   As stated before, such subtractions are not visible through the Fourier
filter at $r >0$ and are thus irrelevant for the determination of $F(r)$.

   At ${\cal O}(p^4)$, the subtracted scalar form factor of the pion reads
\beq
\label{FS}
  \bar F_S(q^2) = 1 - \frac{m_\pi^2}{16\pi^2 F_\pi^2}
     \left[ \frac{2x-1}{2} J^{(0)}(x) + \frac{19x^2-10x}{120} \right],
\eeq
   where $x=q^2/m_\pi^2$, and the function $J^{(0)}(x)$ is defined
as\footnote{The results for $0\leq x<4$ and $4<x$ are obtained by
analytical continuation.}
\beq
\label{J0}
     J^{(0)}(x) = \int_0^1 \ln [1+x(y^2-y)-i0^+] \,dy
   ~=~  -2 - \sigma \ln\Big(\frac{\sigma-1}{\sigma+1}\Big), \quad
      \sigma \equiv \sqrt{1 - \frac{4}{x}},\quad x<0.
\eeq
   As numerical values, we use $F_\pi=92.4$ MeV \cite{PDG00} and the
charged pion mass $m_\pi=139.6$ MeV.
   The polynomial in Eq.\ (\ref{FS}) results in vanishing first and
second derivatives of $\bar{F}_S$ at $t=0$.

   At one-loop order, the subtracted vector form factor is of a similar form:
\beq
\label{FV}
  \bar F_V(q^2) = 1 - \frac{m_\pi^2}{16\pi^2 F_\pi^2}
     \left( \frac{x-4}{6} J^{(0)}(x) + \frac{3x^2-20x}{180} \right).
\eeq
   At the two-loop level, the scalar and vector form factors are given by
more lengthy expressions which can be found in Refs.\
\cite{colangelo96,bijnens98}.
   To be specific, we made use of Eqs.\ (3.6)--(3.8) and
(3.16)--(3.18) of Ref.\ \cite{bijnens98}, using the parameters
(low-energy constants) $\bar l_1=-1.7$, $\bar l_2=6.1$, $\bar l_3=2.9$,
$\bar l_4=4.472$, $\bar l_6=16.0$ (set I \cite{bijnens98}).

   For comparison, we discuss as another example the isovector
electric form factor of the nucleon, $G_E^V(q^2)$, to leading
nontrivial order [$\O(p^3)$] within HBChPT \cite{bernard92a,fearing97}.
With the above two subtractions one obtains%
\footnote
{In this case one subtraction would be sufficient to remove
low-energy constants.  Note that the
functions $J(q^2)$ and $\zeta(q^2)$ used in Ref.\ \cite{bernard92a} are
related with the function $J^{(0)}(x)$, Eq.\ (\ref{J0}), by
$16\pi^2 \zeta(q^2) = -J^{(0)}(x)$ and
$(96\pi^2/m_\pi^2) J(q^2) = -2x/3 + (x-4)J^{(0)}(x)$.}
\beqn
\label{FV-N}
  \bar G_E^V(q^2) = 1 &-& \frac{m_\pi^2}{16\pi^2 F_\pi^2}
     \left( \frac{x-4}{6} J^{(0)}(x) + \frac{3x^2-20x}{180} \right)
\nn
         &-& \frac{m_\pi^2 g_A^2}{16\pi^2 F_\pi^2}
     \left( \frac{5x-8}{6} J^{(0)}(x) + \frac{21x^2-40x}{180} \right),
\eeqn
where $g_A=1.267$ is the axial-vector coupling constant.

   The scalar form factor of the nucleon and the corresponding spatial
distribution were recently discussed by Robilotta \cite{robilotta01} in
the context of the two-pion-exchange contribution to the $NN$ potential
(see Fig.\ 8 of that reference).

\begin{figure}[tbp]
\centerline{
\includegraphics[height=6.5cm]{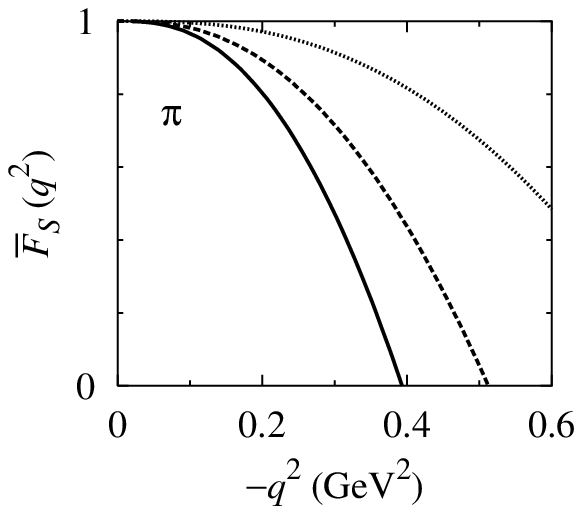}
\includegraphics[height=6.5cm]{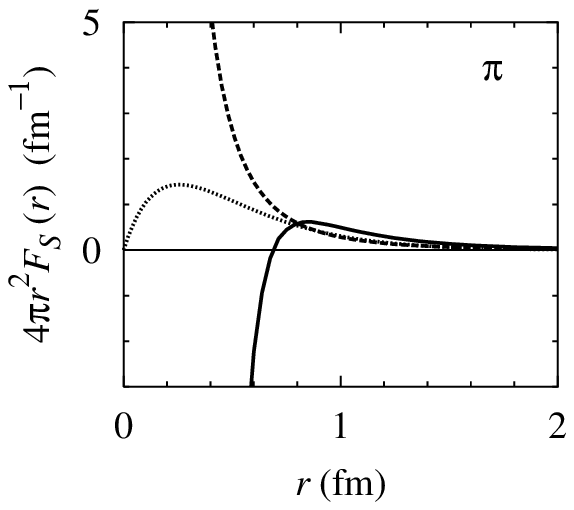}
}
\centerline{
\includegraphics[height=6.5cm]{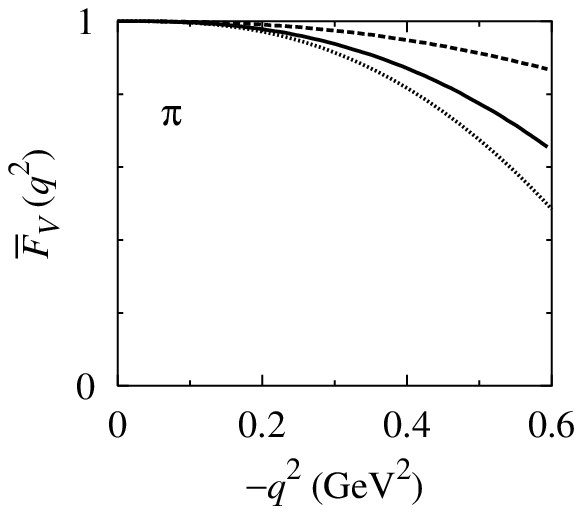}
\includegraphics[height=6.5cm]{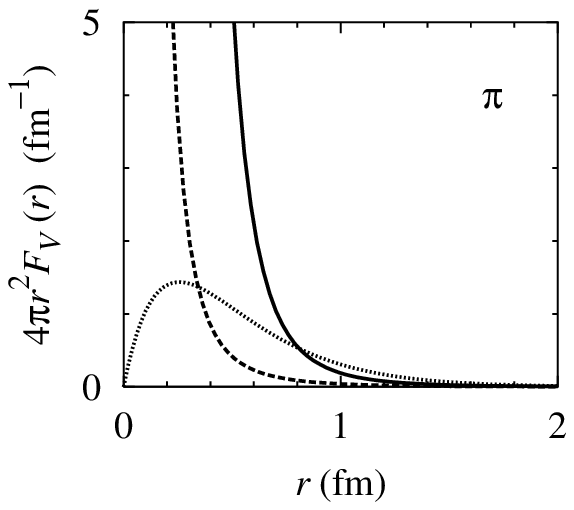}
}
\centerline{
\includegraphics[height=6.5cm]{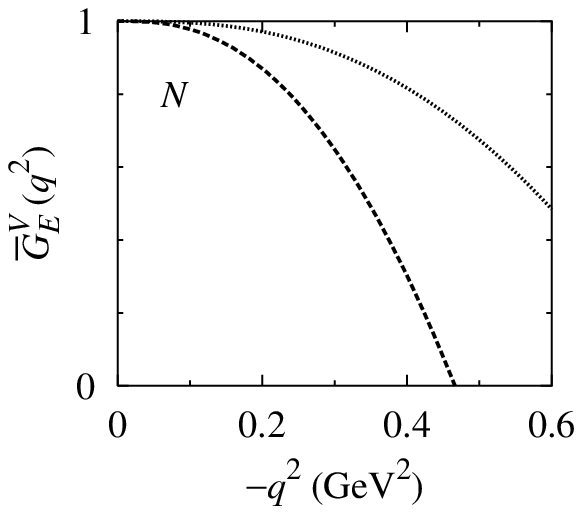}
\includegraphics[height=6.5cm]{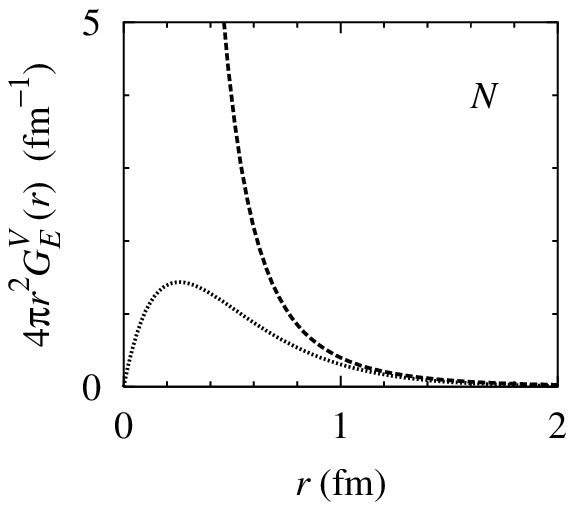}
}
\bigskip
\caption{Left panels: the (twice-subtracted) scalar and vector form
factors of the pion in ChPT at order $\O(p^6)$
\protect\cite{bijnens98} and the isovector charge form
factor of the nucleon in HBChPT at order $\O(p^3)$
\protect\cite{bernard92a,fearing97}.
Right panels: corresponding spatial distributions obtained as Fourier
transforms.
Dashed lines: one-loop prediction.
Solid  lines: two-loop prediction (for pions only).
Dotted lines: prediction of the pole dominance model due to a scalar
or vector meson of mass 770 MeV.}
\label{fig:FF}
\end{figure}

   In Fig.~\ref{fig:FF} we show the three form factors $\bar{F}(q^2)$ of
Eqs.\ (\ref{FS}), (\ref{FV}), and (\ref{FV-N}) and their corresponding
Fourier transforms $F(r)$ which, in the case of $F=F_V$ and $F=G_E^V$,
are interpreted as the electric charge density.\footnote
{Analytical representations of $F(r)$ in ChPT at one-loop order are given
in Appendix~\ref{appendix}.}
   For the sake of comparison we also present the results of a
pole approximation to those form factors,
\beq
   F(q^2) = \frac{m^2}{m^2 - q^2}, \qquad
   4\pi r^2 F(r) = m^2\, r\, e^{-mr},
\eeq
with the mass scale $m=770$ MeV arbitrarily chosen to be the
$\rho$-meson mass.

   As a first observation, we note that all spatial distributions considered
are positive at the one-loop order.
   Such a positive sign confirms a (maybe too) naive understanding of the
peripheral structure of the target as being created by a cloud of virtual
pions
which create, in the case of a $\pi^+$ or proton target, a positive electric
charge density of finite size.
   Also the scalar density seen by an external scalar field carries
a positive sign.
   The QCD coupling of an external scalar field $\tilde{s}(x)$ to the
scalar density $S(x)$,
\beq
\label{lextscalar}
   {\cal L}_{\rm ext}=-\hat{m}[\bar{u}(x)u(x)+\bar{d}(x)d(x)]\tilde{s}(x),
\eeq
   is described through the Lagrangian
$\frac14 F^2 {\rm Tr}\,(\chi U^\dagger + U\chi^\dagger)$
at lowest order in ChPT,
where, in the present case, $\chi=m_\pi^2 \tilde{s}(x)$.
   Inserting $U=(\sigma+i\vectau\cdot\vecpi)/F$, the lowest-order interaction
reads
\beq
   {\cal L}_{\rm ext}^{\rm eff}=-\frac12 m_\pi^2 \vecpi^2(x) \tilde{s}(x)
       \equiv -S^{\rm eff}(x)\tilde{s}(x).
\eeq
   Thus, $S^{\rm eff}(x)$ has manifestly positive
matrix elements which, when probed through $\tilde{s}$
as part of the pion cloud, lead to a positive density.

   When two-loop corrections are taken into account, the spatial distributions
at small distances, $r \alt 1$ fm, change drastically.
   In particular, the charge density $F_V(r)$ of the pion which at the
one-loop level was concentrated at $r \alt 0.3$ fm, now extends up to
$r \sim 0.6$ fm.
   An even more dramatic effect is observed for the scalar density $F_S(r)$,
which due to the two-loop contribution changes sign at $r \alt 0.7$ fm.
   However, one should keep in mind that ChPT is a low-energy effective field
theory and it is likely that higher-order corrections will change the spatial
distributions at short distances substantially.
   An indication for such a scenario is given by the dotted
lines in Fig.~\ref{fig:FF} which refer to short-range mechanisms
such as vector mesons.
   Of course, such mechanisms are only included in ChPT by means of
low-energy couplings in the effective Lagrangian.

   Finally, let us emphasize again that the above examples
do not pretend to {\em prove} that spatial distributions $F(r)$ have a strict
operational sense.
   However, they certainly suggest a simple intuitive picture of the hadron
periphery which is the natural domain of ChPT.

\subsection{Polarizabilities}

   We now extend the discussion to spatial distributions associated
with the generalized polarizabilities as introduced in Sec.\ \ref{sec:LEX}.
   The predictions for the generalized dipole polarizabilities
$\alpha_L(q^2)$ and $\beta(q^2)$ of the nucleon in the framework
of SU(2)$_f$ HBChPT at $\O(p^3)$ read \cite{hemmert00,lvov94}
\beqn
\label{alphasN}
     \alpha_L(q^2) &=& E_\pi^2 \int_0^1
       [8s^2 + s m_\pi^2 (16y^2-16y+9) - 3 m_\pi^4 (2y-1)^2]
       \, \frac{dy}{32s^{5/2}}
\nn
       && \qquad = \frac{E_\pi^2}{8m_\pi a(a+1)}
       \Big[(a+1)(2a-1)\frac{{\rm arctan}\,\sqrt a}{\sqrt a} +2a+1 \Big],
\nn
     \beta(q^2) &=& E_\pi^2 \int_0^1
       (2y-1)^2 (s+m_\pi^2) (4s-3m_\pi^2)\,\frac{dy}{16s^{5/2}}
\nn
       && \qquad = \frac{E_\pi^2}{16m_\pi a(a+1)}
       \Big[(a+1)(2a+1)\frac{{\rm arctan}\,\sqrt a}{\sqrt a} -2a-1 \Big],
\eeqn
where $s = m_\pi^2 + q^2(y^2-y)$, $a=-q^2/(4m_\pi^2)$,
and $E_\pi = eg_A\sqrt{2}/(8\pi F_\pi)$ is the Kroll-Ruderman amplitude
of $\pi^\pm$ photoproduction in the chiral limit.\footnote
{The integral representations of Eq.\ (\ref{alphasN})
are, of course, less convenient than the equivalent elementary formulas.
   They are given here only as a
historical reference, in the form in which they were first
calculated and reported \cite{lvov94}.
   The same results were found independently by the authors of
Refs.\ \cite{metz96,hemmert97,hemmert00}.
   Recently, results have also been given in the framework of the
Small Scale Expansion including the $\Delta$-isobar as a
dynamic degree of freedom \cite{hemmert00}.}

The transverse electric polarizability of the nucleon remains yet to be
determined.

   There are no published calculations of the nucleon's generalized
polarizabilities $\alpha_L(q^2)$ and $\beta(q^2)$ within SU(3)$_f$
HBChPT, except for the case $q^2=0$ considered in Refs.\
\cite{bernard92,butler92}.
   However, it is straightforward to extend the results of Eq.\ (\ref{alphasN})
to the SU(3)$_f$ case in the limit of equal $N$, $\Lambda$, and $\Sigma$
masses.
   In this limit, the {\em square} baryon-octet mass differences is
considered as small in comparison with the square kaon mass, $m_K^2$,
which empirically is a good approximation.
   Then, the structure of Feynman diagrams contributing to Compton scattering
in the SU(3)$_f$ and SU(2)$_f$ cases, respectively, is very similar, the only
difference resulting (a) from different meson-baryon couplings for
charged mesons (or, stated differently, from different Kroll-Ruderman
amplitudes) and (b) from different masses of the Goldstone bosons
entering the appropriate loop diagrams.
   The generalized polarizabilities of the nucleon in SU(3)$_f$ HBChPT
receive, in addition to the results of Eqs.\ (\ref{alphasN}),
contributions due to kaon loops which are given by the same expressions
as in Eqs.\ (\ref{alphasN}) after the replacements $m_\pi\to m_K$ and
$E_\pi \to E_K$, where
\beqn
    E_\pi^2 &\equiv& E_{\rm KR}^2(\gamma N\to \pi^\pm N) =
            \Big(\frac{e\sqrt 2}{8\pi F_\pi}\Big)^2 I_\pi \, ,
\nn
    E_K^2 &\equiv&
       E_{\rm KR}^2(\gamma N\to K^+ \Lambda)
     + E_{\rm KR}^2(\gamma N\to K^+ \Sigma) =
            \Big(\frac{e\sqrt 2}{8\pi F_K}\Big)^2 I_K
\eeqn
and
\beq
     I_\pi= g_A^2 = (D+F)^2, \qquad
     I_K = \cases{ \frac23 D^2 + 2F^2, & proton  \cr
                          (D-F)^2,     & neutron }
\eeq
(cf.\ \cite{bernard92,butler92}).
   In an $\O(p^3)$ calculation, the difference between the pion and
kaon decay constants is of higher order.
   Empirically, $F_K\simeq 1.22 F_\pi$ \cite{PDG00},
but in our numerical analysis we make use of a universal value
$F_K = F_\pi$ with $F_\pi=92.4$ MeV.
   Furthermore, we insert the SU(3)$_f$ ratio $F/D=2/3$.

   By applying the same procedure to the other octet baryons,
one obtains a generic polarizability $F(q^2)$ as
\beq
   F(q^2) =
     \Big[ {\rm Eq.}~(\ref{alphasN}) \Big]
        \times \frac{I_\pi}{g_A^2}
  ~+~ \Big[ {\rm Eq.}~(\ref{alphasN}) \Big]_{m_\pi\to m_K}
        \times \frac{I_K}{g_A^2}.
\eeq
   The SU(3)$_f$ coefficients $I_\pi$ and $I_K$ are identical with
the ones given in Eq.~(4) of Ref.\ \cite{bernard92}
for the case $q^2=0$.
   For convenience, we collect
these coefficients in Table~\ref{tab:Ipi-IK} (using $F/D=2/3$).
   Note, however, that the pion-loop contribution for $\Lambda$ and
$\Sigma^\pm$ is perhaps not given very accurately by this procedure,
since the baryon mass difference in the transitions $\pi^\pm\Lambda
\leftrightarrow \Sigma^\pm$ is not fully negligible in comparison with
the pion mass $m_\pi$.

\begin{table}[htb]
\caption{Flavor coefficients $I_\pi$ and $I_K$ determining pion and
kaon loop contributions to the generalized polarizabilities of octet
baryons in SU(3)$_f$ HBChPT at order $\O(p^3)$.}
\label{tab:Ipi-IK}
\begin{tabular}{ccccc}
  $B$        & $I_\pi$     & $I_\pi/g_A^2$ & $I_K$         & $I_K/g_A^2$ \\
\hline
  $p$        & $(D+F)^2$           &  1.00 & $\frac23 D^2 +2F^2$ &  0.56 \\
  $n$        & $(D+F)^2$           &  1.00 & $(D-F)^2$           &  0.04 \\
  $\Lambda$  & $\frac43 D^2$       &  0.48 & $\frac13 D^2 +3F^2$ &  0.60 \\
  $\Sigma^+$ & $\frac23 D^2 +2F^2$ &  0.56 & $(D+F)^2$           &  1.00 \\
  $\Sigma^0$ & $ 4F^2$             &  0.64 & $ D^2+F^2$          &  0.52 \\
  $\Sigma^-$ & $\frac23 D^2 +2F^2$ &  0.56 & $(D-F)^2$           &  0.04 \\
  $\Xi^0$    & $(D-F)^2$           &  0.04 & $(D+F)^2$           &  1.00 \\
  $\Xi^-$    & $(D-F)^2$           &  0.04 & $\frac23 D^2 +2F^2$ &  0.56
\end{tabular}
\end{table}

   At $\O(p^4)$, SU(2)$_f$ and SU(3)$_f$ ChPT predictions for
all three dipole polarizabilities $\alpha_L(q^2)$, $\alpha_T(q^2)$, and
$\beta(q^2)$ have been reported for pions and kaons \cite{unkmeir00,fuchs00}.
   At that order, the three generalized dipole polarizabilities are
degenerate, i.e.
\beq
   \alpha_L(q^2) = \alpha_T(q^2) = -\beta(q^2)
   \quad\mbox{and}\quad
   \alpha_L(r) = \alpha_T(r) = -\beta(r),
\eeq
and hence we only need to discuss, say, the generalized magnetic
polarizability which can be expressed as \cite{fuchs00}
\beq
\label{betasM}
    -\beta(q^2)
      = \frac{e^2}{4\pi M}\,\frac{1}{(4\pi F_\pi)^2}\,
   \Big[ A + (B_\pi + C_\pi x_\pi) {J^{(0)}}'(x_\pi)
        + C_K x_K {J^{(0)}}'(x_K) \Big].
\eeq
Here, $M$ is the mass of the particle in question,
\beq
     x_\pi   =\frac{q^2}{m_\pi^2} \, , \quad
     x_K =\frac{q^2}{m_K^2} \, ,
\eeq
the function ${J^{(0)}}'(x)$ is given by
\beq
\label{J0prime}
   {J^{(0)}}'(x) =
        \frac{J^{(0)}(x)}{dx} = \frac{2J^{(0)}(x) + x}{x(x-4)},
\eeq
and $A$, $B_\pi$, $C_\pi$, and $C_K$ are constants given below.
   The terms in Eq.\ (\ref{betasM}) depending on $x_\pi$ and $x_K$ represent
contributions of pion and kaon loops, respectively.
   The $q^2$-independent term proportional to $A$ originates from a
contribution at short distances represented by low-energy constants
in the effective chiral Lagrangian
\cite{gasser85}:
\beqn
    A(\pi^\pm) &=& A(K^\pm) = 64\pi^2 (L_9^r + L_{10}^r)
          = \frac{2F_A}{F_V} = 0.90 \pm 0.12,
\nn
    A(\pi^0) &=& A(K^0) = A(\bar K^0) =0.
\eeqn
   The numerical value of $A(\pi^\pm)$ is fixed \cite{terentev72}
by the experimentally known axial ($F_A$)
and theoretically known vector ($F_V$) form factors
of the radiative pion decay $\pi^+ \to e^+\nu_e\gamma$
using $F_A/F_V= 0.448 \pm 0.062$ \cite{PDG00}.
   The relation $A(\pi) = A(K)$ is valid in the SU(3)$_f$-symmetry limit.
   The other constants entering Eq.\ (\ref{betasM}) are flavor-dependent
coefficients which determine contributions of pion and kaon loops,
respectively:
\beq
   C_\pi  = \cases{-1   & for $\pi^0$,   \cr
                   -1/2 & for $\pi^\pm$, \cr
                   -1/4 & for kaons,}
\quad
   B_\pi  = \cases{ 1   & for $\pi^0$,   \cr
                    0   & otherwise,}
\quad
   C_K    = \cases{-1/2 & for $K^\pm$,   \cr
                   -1/4 & otherwise.}
\eeq

\begin{figure}[tbp]
\centerline{
\includegraphics[height=6.5cm]{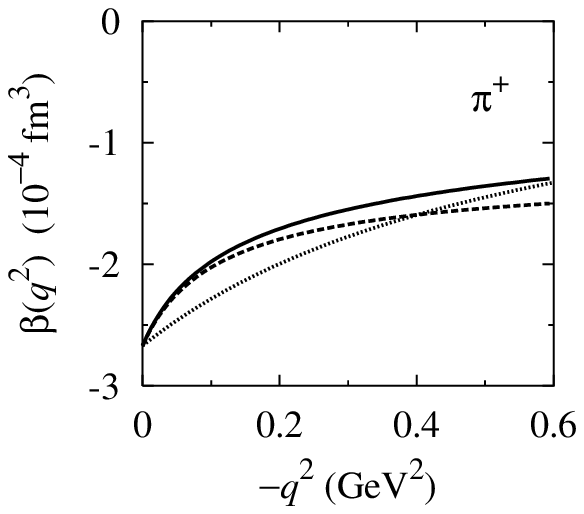}
\includegraphics[height=6.5cm]{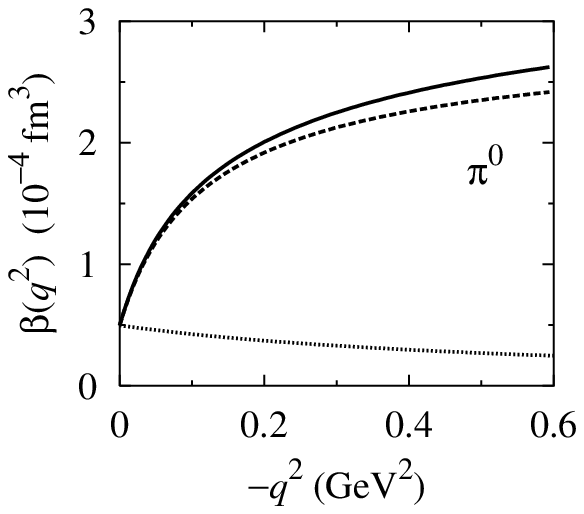}
}
\centerline{
\includegraphics[height=6.5cm]{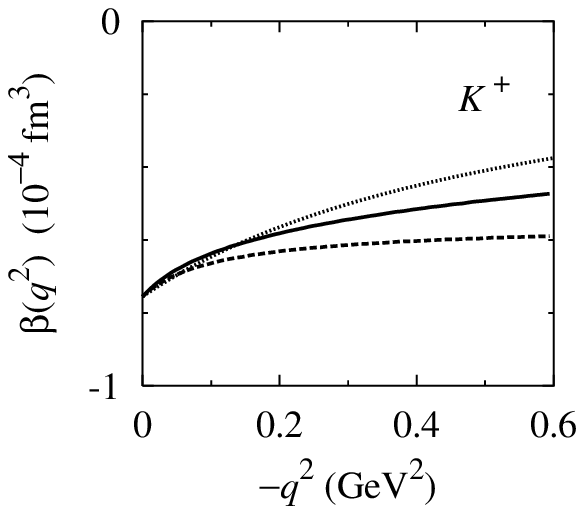}
\includegraphics[height=6.5cm]{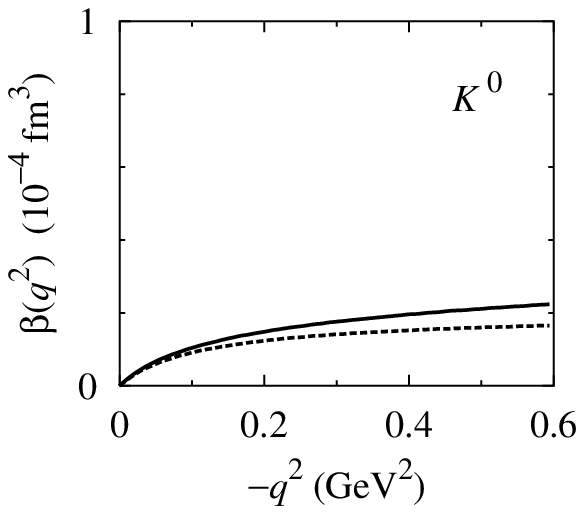}
}
\bigskip
\caption{Generalized magnetic polarizability $\beta(q^2)$ of pions and
kaons at $\O(p^4)$ \protect\cite{unkmeir00,fuchs00}.
Dashed lines: contribution of pion loops.
Solid  lines: contribution of pion and kaon loops.
Dotted lines: VMD predictions normalized to $\bar\beta$
as given by SU(3)$_f$ ChPT.}
\label{fig:q-beta-mesons}
\end{figure}

\begin{figure}[tbp]
\centerline{
\includegraphics[height=6.5cm]{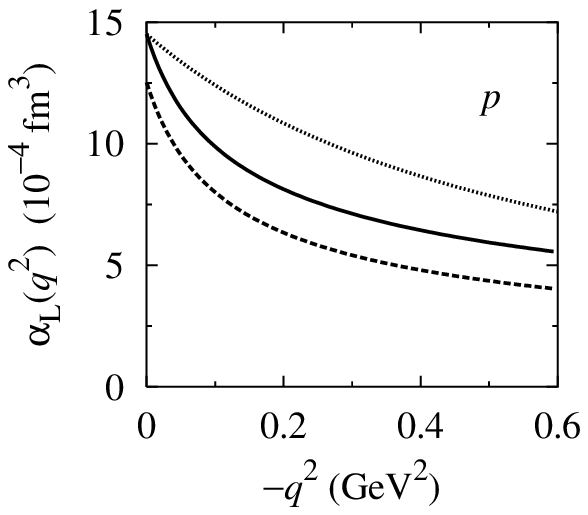}
\includegraphics[height=6.5cm]{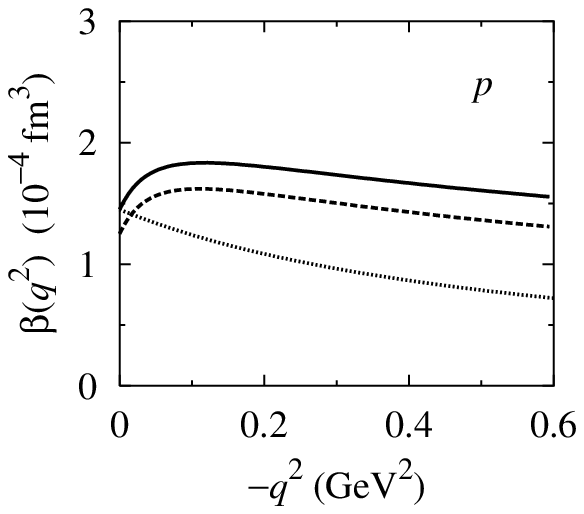}
}
\centerline{
\includegraphics[height=6.5cm]{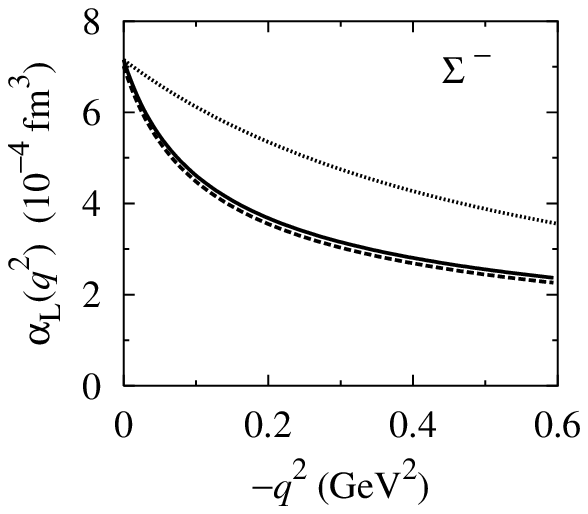}
\includegraphics[height=6.5cm]{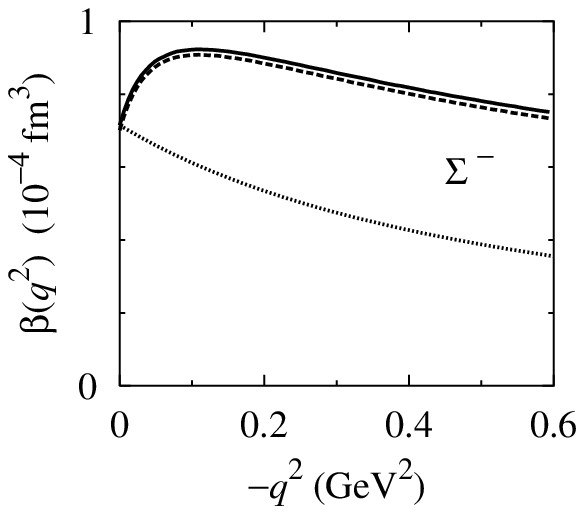}
}
\centerline{
\includegraphics[height=6.5cm]{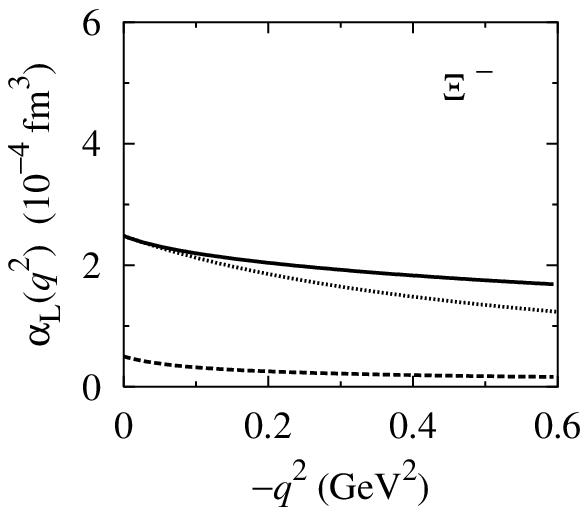}
\includegraphics[height=6.5cm]{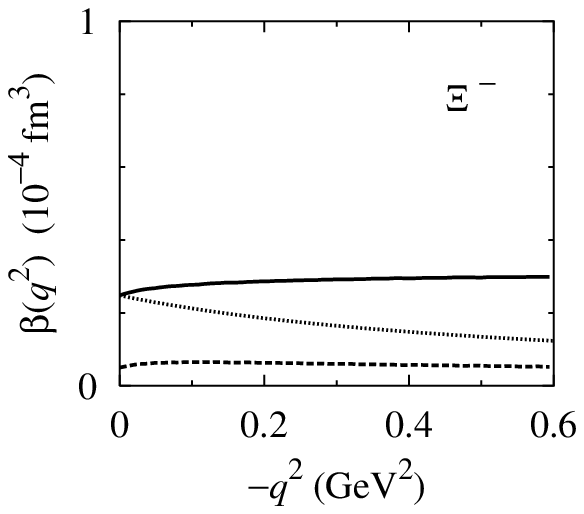}
}
\bigskip
\caption{Generalized longitudinal electric and magnetic
polarizabilities $\alpha_L(q^2)$ and $\beta(q^2)$ of the proton,
the $\Sigma^-$, and the $\Xi^-$ at $\O(p^3)$
(see \protect\cite{hemmert00,lvov94} and the text).
Dashed lines: contribution of pion loops.
Solid  lines: contribution of pion and kaon loops.
Dotted lines: VMD predictions normalized to $\bar\alpha$ and $\bar\beta$
as given by SU(3)$_f$ ChPT.}
\label{fig:q-beta-baryons}
\end{figure}

   The generalized polarizability of the Goldstone bosons are shown
together with the results for the proton, the $\Sigma^-$, and the $\Xi^-$
in Figs.~\ref{fig:q-beta-mesons} and \ref{fig:q-beta-baryons}.

\begin{figure}[tbp]
\centerline{
\includegraphics[height=6.5cm]{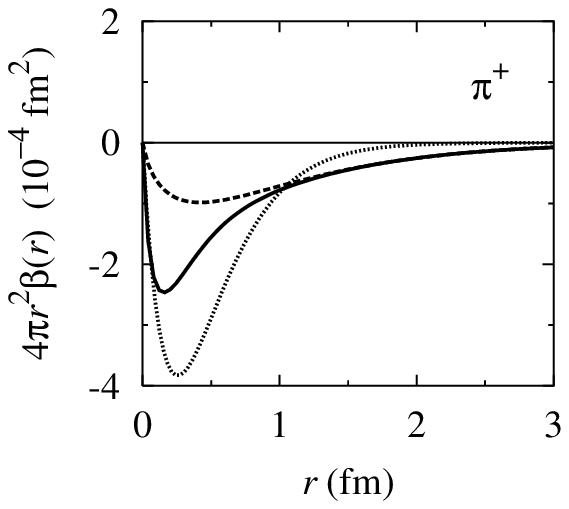}
\includegraphics[height=6.5cm]{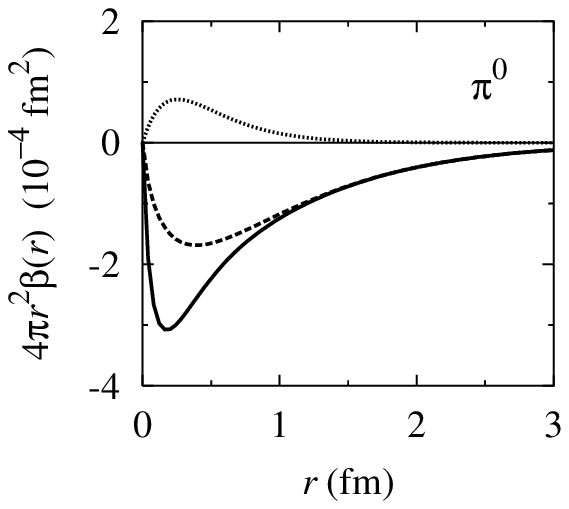}
}
\centerline{
\includegraphics[height=6.5cm]{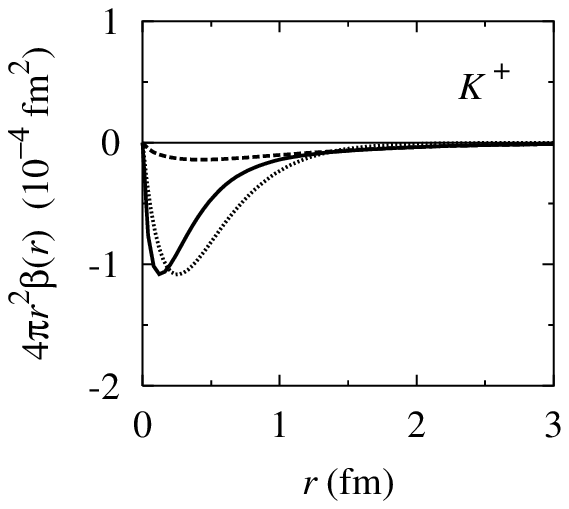}
\includegraphics[height=6.5cm]{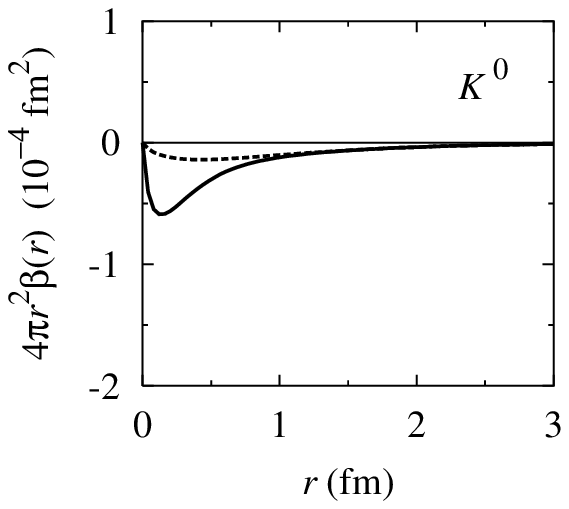}
}
\bigskip
\caption{Density of the magnetic polarizability $\beta(r)$ of pions
and kaons at $\O(p^4)$.
The $\delta$-singularity at $r=0$ is not shown
(see the discussion of Eq.\ (\protect\ref{betas_over_r>0}) in the text).
Dashed lines: contribution of pion loops.
Solid  lines: contribution of pion and kaon loops.
Dotted lines: VMD predictions normalized to $\bar\beta$
as given by SU(3)$_f$ ChPT.}
\label{fig:r-beta-mesons}
\end{figure}

\begin{figure}[tbp]
\centerline{
\includegraphics[height=6.5cm]{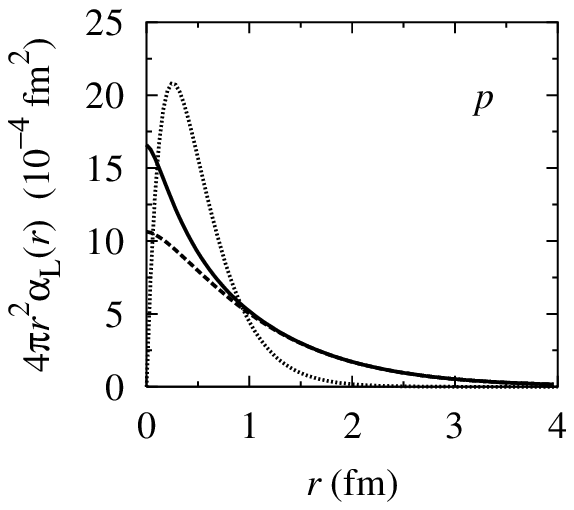}
\includegraphics[height=6.5cm]{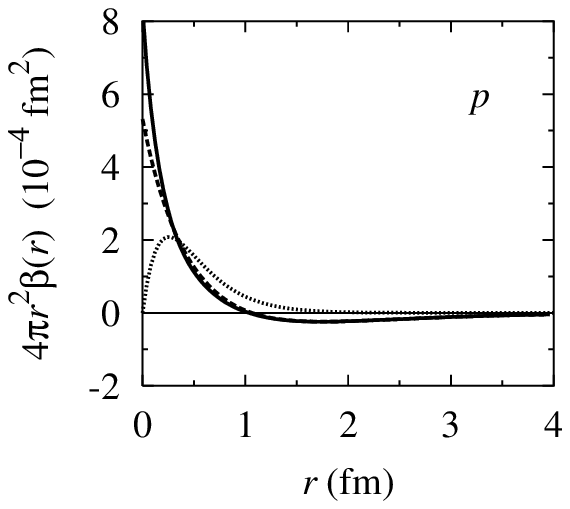}
}
\centerline{
\includegraphics[height=6.5cm]{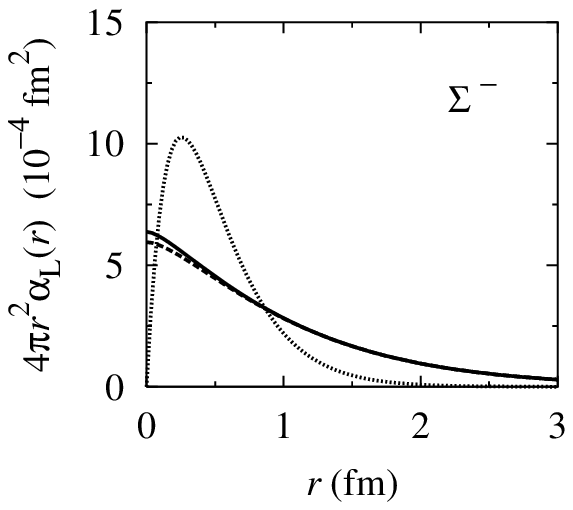}
\includegraphics[height=6.5cm]{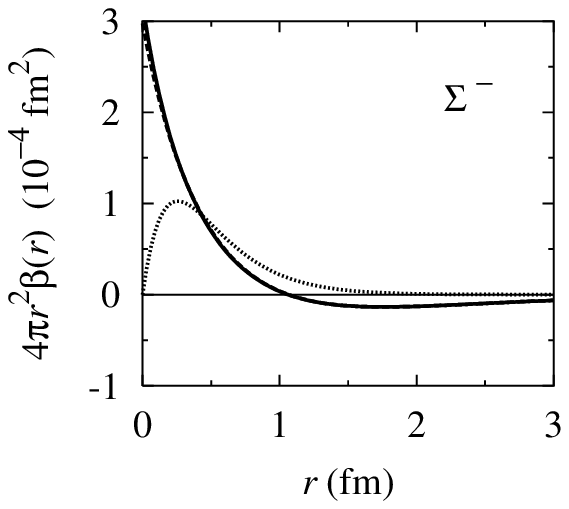}
}
\centerline{
\includegraphics[height=6.5cm]{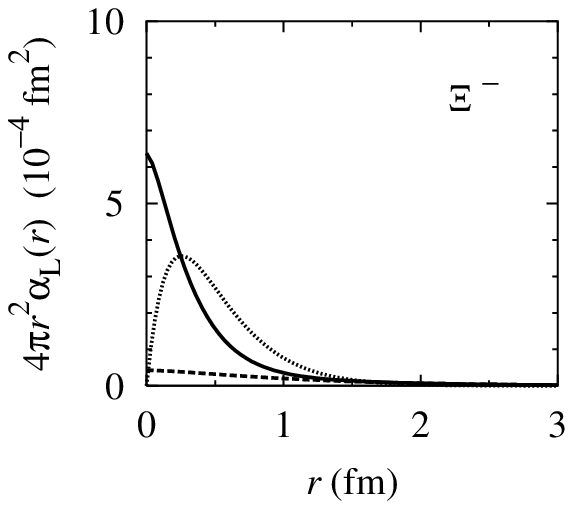}
\includegraphics[height=6.5cm]{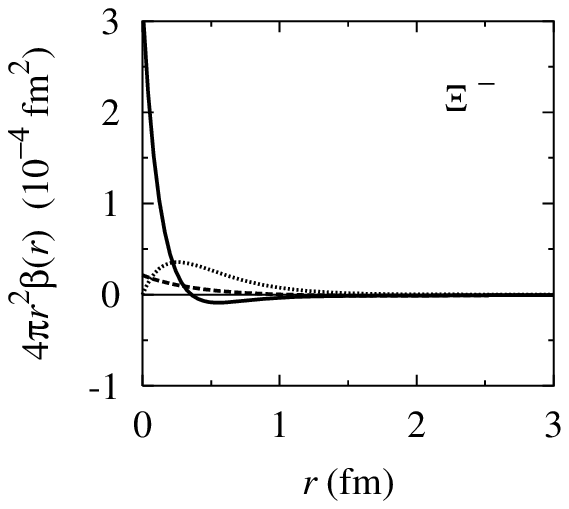}
}
\bigskip
\caption{Density of the longitudinal electric and magnetic polarizabilities
$\alpha_L(r)$ and $\beta(r)$ of the proton, the $\Sigma^-$, and the $\Xi^-$
at $\O(p^3)$.
Dashed lines: contribution of pion loops.
Solid  lines: contribution of pion and kaon loops.
Dotted lines: VMD predictions normalized to $\bar\alpha$ and  $\bar\beta$
as given by SU(3)$_f$ ChPT.}
\label{fig:r-beta-baryons}
\end{figure}

   The spatial distributions calculated as the Fourier transforms of the
generalized polarizabilities are shown in
Figs.~\ref{fig:r-beta-mesons} and \ref{fig:r-beta-baryons}.
   The corresponding analytical results obtained through Eq.~(\ref{Yukawa})
are given in Appendix \ref{appendix}.
   For large $r$, the pion and kaon loop contributions to a generic
polarizability $F(r)$ follow an exponential behavior
$e^{-2m_\pi r}$ and $e^{-2m_K r}$,
respectively, as determined by the nearest singularities of
$F(q^2)$ for time-like momenta, $q^2=4m_\pi^2$ and $q^2=4m_K^2$.
   The $\delta$-singularity at $r=0$ cannot be seen in these plots.
   However, at least for mesons, such a singularity exists for sure within
ChPT, and it is determined by the asymptotic value of the polarizability for
$q^2\to -\infty$.
   Thus, the integrals of the spatial distributions over $r > 0$
are
\beq
\label{betas_over_r>0}
     \lim_{\epsilon\to 0+} \int_{\epsilon}^\infty 4\pi r^2 \beta(r)\,dr
   ~=~ \beta(q^2=0) - \beta(q^2=-\infty).
\eeq
    The generalized polarizabilities of the octet baryons, given by Eqs.\
(\ref{alphasN}) and (\ref{betasM}), vanish at infinity,
so that the integral (\ref{betas_over_r>0}) gives just $\bar\beta$
in this case.
   This is not the case for mesons, since $x{J^{(0)}}'(x) \to 1$
for $x\to -\infty$.
   The corresponding $\delta$-singularity in $4\pi r^2 \beta(r)$
is driven by the limits \cite{fuchs99}
\beqn
\label{asmp_pol}
   \lim_{q^2\to-\infty}\beta_{\pi^\pm}(q^2)&=&\bar{\beta}_{\pi^\pm}
      +\frac{9}{2}\bar{\beta}_{\pi^0},
\nn
   \lim_{q^2\to-\infty}\beta_{\pi^0}(q^2)
        &=&\frac{15}{2}\bar{\beta}_{\pi^0},
\nn
   \lim_{q^2\to-\infty}\beta_{K^\pm}(q^2)&=&\bar{\beta}_{K^\pm}
       +\frac{9}{2}\frac{m_\pi}{m_K}\bar{\beta}_{\pi^0},
\nn
   \lim_{q^2\to-\infty}\beta_{K^0}(q^2)&=&
       3\frac{m_\pi}{m_K}\bar{\beta}_{\pi^0}.
\eeqn
   The SU(2)$_f$ results for pions are found in Eqs.\ (39)
and (40) of Ref.\ \cite{unkmeir00}.

   It is very interesting that the loop contributions behave exactly as
one would expect from a classical interpretation of the Langevin diamagnetism.
   In such a picture, a change in the external magnetic field would produce
circulating currents induced in the charge density of the meson cloud
which on their part give rise to an induced magnetization.
   Both pion and kaon clouds are seen to be diamagnetic, at least at large
distances $r\gtrsim 1/m_\pi$.
   Simultaneously, these clouds generate a positive sign for the electric
polarizability.
   However, in the case of the nucleon the diamagnetic character of the pion
cloud  disappears at distances $r\lesssim 1/m_\pi$, where paramagnetism
prevails and $\beta(r)$ becomes positive.

   When a hadron is probed by photons of very small (space-like) momenta
$|\vecq| \ll m_\pi$, the magnetic response is essentially only sensitive
to $\beta(q^2=0)=\bar\beta$.
   When the momentum $|\vecq|$ increases, the very peripheral,
negative part of the magnetic polarizability of the pion cloud no longer
contributes to
\beq
\label{betaq2}
     \beta(q^2)=\int \beta(r) \exp(i\vecq\cdot\vecr) \,d\vecr,
\eeq
due to the oscillatory behavior of the integrand for large distances.
   This explains why the magnetic polarizability
$\beta(q^2)$ for {\em all} the particles (pions, kaons, nucleons, etc.)
universally gets a positive increase (see
Figs.~\ref{fig:q-beta-mesons} and \ref{fig:q-beta-baryons}).
   The slope of $\beta(q^2)$ at the origin is proportional to the mean
square radius of the spatial distribution,
\beq
\label{beta-slope}
   \frac{d\beta(0)}{dq^2} = \frac{2\pi}{3} \int_0^\infty r^4 \beta(r)\,dr.
\eeq
   Obviously, for all mesons, the cloud distribution $4\pi r^2 \beta(r)\leq 0$
for all $r\geq 0$. Hence, the slope is negative as a function of $q^2$
(positive, when plotted against $-q^2$).
   Also, the curvature as a function of $q^2$ is concave.
   For the proton, the $\Sigma^-$, and the $\Xi^-$
the asymptotic negative pion and kaon tails in the integral
(\ref{beta-slope}) dominate over the positive contribution coming from
distances smaller than 1 fm, 1 fm, and 0.4 fm, respectively.
   This makes the slopes of the magnetic polarizability of all baryons
considered positive as a function of $-q^2$ as well.
   On the other hand, such behavior is in some cases opposite to that
expected from VMD.
   Suppose the $q^2$ dependence of $\beta(q^2)$ was determined by the $\rho$
or $\omega$ mesons mediating electromagnetic interactions,
\beq
      \Big[\beta(q^2)\Big]_{\rm VMD} =
        \bar\beta\,\frac{m_\rho^2}{m_\rho^2 - q^2}.
\eeq
   From this, the spatial distribution of $\beta(r)$
\beq
      \Big[4\pi r^2 \beta(r)\Big]_{\rm VMD} =
        m_\rho^2 \, r \, e^{-m_\rho r}\,\bar\beta
\eeq
   would have the same the sign as $\bar\beta$ for all $r>0$.
   For the $\pi^0$ and for the octet baryons, having positive
$\bar\beta$, such a sign is in conflict with the ChPT behavior
[see Figs.~\ref{fig:q-beta-baryons} and~\ref{fig:r-beta-baryons}
in which the VMD distributions are also shown].
   From a phenomenological point of view, the full magnetic polarizability
$\bar\beta$ of the neutral pion should be approximately three times the
$\O(p^4)$ prediction, mainly due to a paramagnetic contribution
of the M1 transitions $\pi^0\to\omega$ or $\rho^0$.
   In other words, a more realistic VMD curve should be three times higher
than that shown in Fig.~\ref{fig:r-beta-mesons},
so that the above conflict with ChPT would be even more severe.

   In the case of the (longitudinal) electric polarizability of
octet baryons the long-range, peripheral part is suppressed with
increasing $-q^2$.
   This part is relatively large for all baryons except for the
$\Xi$'s, and thus $\alpha_L(q^2)$ shows a rapid
decrease with increasing space-like $q$
which is steeper than for VMD [see Fig.~\ref{fig:q-beta-baryons}].

   Another instructive feature of the plots shown in
Figs.~\ref{fig:r-beta-mesons} and \ref{fig:r-beta-baryons}
is that the kaon-loop contribution is concentrated at about the same scale of
$r$ (or even at smaller scales) as the VMD contribution
due to the large mass of kaon pairs, $2M_K \approx 1$ GeV.
   Kaon-loop contributions exactly fall into the short-range region where
one finds other contributions of similar range treated via low-energy
constants of the effective Lagrangian of ChPT.
   However, at ${\cal O}(p^3)$ no such counter terms contribute to the
generalized polarizabilities.
   In other words, even though the relevant counter terms are formally of
higher order in the power counting of ChPT, one may wonder whether their
quantitative importance is underrated as compared to the kaon-loop
contributions.
   Therefore, quantitative conclusions drawn from calculations keeping kaon
loops and ignoring short-range contributions have to be treated with some
care.
   For a similar conclusion, see Ref.~\cite{donoghue99}.

   The spatial distribution of $\beta(r)$ for the nucleon shown in
Fig.~\ref{fig:r-beta-baryons} may also shed light on an old question
which has remained open for more than 30 years:
   Why is the magnetic polarizability $\bar\beta$ of the proton
so small ($\bar\beta_p \approx 2\times 10^{-4}~\rm fm^3$ \cite{PDG00}),
despite of a very large paramagnetic contribution of the
$\Delta$ resonance which, in various evaluations based on quark models,
dispersion theories, effective Lagrangians, etc., ranges between +7 to
$+13\times 10^{-4}~\rm fm^3$ (see, for example, \cite{lvov93,bernard93}).
   In this context, it is sometimes stated that the pion cloud produces
a large diamagnetic (i.e.\ negative) contribution to $\bar\beta$ owing to
the Langevin mechanism producing a negative magnetic susceptibility of bound
charges.
    This point of view became especially popular after calculations of
$\bar\beta_N$ in the Skyrme model (see, for example
\cite{scoccola90,scherer92,gobbi96}),
in which the pion field of the soliton produces
indeed a very big diamagnetic susceptibility resulting from the
pion-cloud tail of the soliton, ranging from $-8$ to
$-16\times 10^{-4}~\rm fm^3$
\cite{scoccola90,scherer92,gobbi96}.
   Figure~\ref{fig:r-beta-baryons}, however, shows that the pion-cloud
periphery with $r \ge 1$~fm carries a very small diamagnetism of only
$-0.45\times 10^{-4}~\rm fm^3$, where the last number is expected to be
a reliable estimate because ChPT, even at leading nontrivial order, should
be reasonable at such distances.
   We conclude that Skyrme models overestimate the pion-cloud contribution
to diamagnetism and that a source for the missing diamagnetism is probably
related with short-range mechanisms rather than with the pion cloud itself.
   In some dispersion theories of Compton
scattering \cite{lvov97}, an additional exchange with a hypothetical
$\sigma$-meson is invoked in order to generate agreement with existing
experimental data.
This, of course, is just an oversimplified model for such a short-range
contribution.

   It would be very interesting and instructive to extend the presently
available ChPT predictions for the generalized polarizabilities of the
nucleon at least to order $\O(p^4)$.
   This would allow one to check whether a modification of the density
$\beta(r)$, including relativistic (nucleon recoil) effects and other
higher-order corrections, indeed provides sufficient diamagnetism
\cite{bernard93}.

\section{Summary and conclusions}

   In the present paper we have developed a covariant formalism
leading to a parametrization of the VCS tensor in terms of
Lorentz invariants free from kinematical singularities and constraints.
   We motivated and performed a gauge-invariant division of the
VCS tensor into contributions of generalized Born terms and
a structure-dependent residual part.
   We then discussed the case of a real final photon and a space-like
virtual initial photon which can be described in terms of three invariant
amplitudes depending on three kinematical variables.
   In the limit $q'\to 0$, the three residual amplitudes reduce to functions of
$q^2$ only which we identified as generalized dipole polarizabilities
$\alpha_L(q^2)$, $\alpha_T(q^2)$, and $\beta(q^2)$.
   All of them can, in principle, be determined in virtual Compton scattering,
although the transverse electric polarizability is not accessible in
experiments sensitive to structure-dependent effects of $\O(q')$ only.
    We proposed a physical interpretation of these polarizabilities
in terms of {\em spatial} distributions of an induced electric polarization
and magnetization, respectively.
    In particular, we argued that a knowledge of all three polarizabilities
is required for a full description of induced polarization phenomena.
   Following this line, we calculated spatial distributions
for pions, kaons, and the baryon octet as Fourier integrals, using the
predictions of ChPT.
   It was found that the distributions obtained confirmed expectations based
on a picture of a hadron's periphery caused by a ``classical" pion cloud.
   Of course, any practical {\em analysis} of experimental data on photon
{\em scattering} will eventually deal with the original, precisely defined
momentum-space form factors, polarizabilities, etc.
   Thus, one should handle the found spatial distributions with care.
   On the other hand, the $r$-space interpretation of such $q$-dependent
quantities clearly allows for a more intuitive visualization in analogy to
the phenomenology and terminology of classical electrodynamics and
nonrelativistic quantum mechanics.

\acknowledgements

A.L.\ thanks the theory group of the Institut f\"ur Kernphysik
for the hospitality and support during his stays in Mainz,
where part of his work was done.
He also thanks the Institute for Nuclear Theory
in Seattle and the organizers of the CEBAF--INT workshop
on ``Probing Nucleon Structure by Real and Virtual Compton Scattering"
(1994) for the hospitality during the time when some of ideas of the present
work were initially elaborated.

\appendix
\section{Spatial distributions to one loop}
\label{appendix}

In this appendix we collect the analytical results for the spatial
distributions in ChPT at the one-loop level which are easily obtained
through Eq.~(\ref{Yukawa}). In the formulas below we use the notation
\beq
     t = q^2,\quad
     x_\pi   = \frac{t}{m_\pi^2},\quad
     x_K = \frac{t}{m_K^2},\quad
     z_\pi   = 2m_\pi r, \quad
     z_K = 2m_K   r.
\eeq

   The imaginary parts of the scalar and vector form factors of the pion,
and of the isovector charge form factor of the nucleon
are determined by the function $J^{(0)}(x)$ which has a branching point
at $x=4$:
\beqn
   \Im F_S(t) &=&  F_S(0)\,\frac{2x_\pi-1}{2} V(x_\pi),
\nn
   \Im F_V(t) &=&  \frac{x_\pi-4}{6} V(x_\pi),
\nn
   \Im G_E^V(t) &=&  \Big(\frac{x_\pi-4}{6}
     + g_A^2 \frac{5x_\pi-8}{6} \Big) V(x_\pi),
\eeqn
where
\beq
     V(x) = \frac{\pi}{(4\pi F_\pi)^2}
           \,\sqrt{\frac{x-4}{x}} \, \theta(x-4).
\eeq
   Evaluating the integral representation of Eq.\ (\ref{Yukawa}) with the
above imaginary parts results in
\beqn
    4\pi r^2 F_S(r) &=& \frac{m_\pi^3 F_S(0)}{(4\pi F_\pi)^2}\,
        \Big[ \frac{48}{z_\pi}\,K_0(z_\pi)
        + \frac{96+14z^2_\pi}{z^2_\pi}\,K_1(z_\pi) \Big],
\nn
    4\pi r^2 F_V(r) &=& \frac{m_\pi^3}{(4\pi F_\pi)^2}\,
        \Big[ \frac{8}{z_\pi}\,K_0(z_\pi) + \frac{16}{z_\pi^2}\,K_1(z_\pi) \Big],
\nn
    4\pi r^2 G_E^V(r) &=& \frac{m_\pi^3}{(4\pi F_\pi)^2}\,
        \Big[ \frac{8+40g_A^2}{z_\pi}\,K_0(z_\pi)
        + \frac{16+(80+8z_\pi^2)g_A^2}{z_\pi^2}\,K_1(z_\pi) \Big],
\eeqn
where $K_\nu(z)$ is the modified Bessel function,
$K_\nu(z) = \int_0^\infty e^{-z\cosh t}\,\cosh(\nu t)\,dt$.

The generalized polarizabilities of the nucleon, Eq.~(\ref{alphasN}),
at time-like momenta $q$ have both a cut starting at $t=4m_\pi^2$ and
a pole at $t=4m_\pi^2$:
\beqn
    && \Im \alpha_L(t) = \frac{\pi}{8}\, E_\pi^2 \left[
       \Big(2 +\frac{4m_\pi^2}{t}\Big)
       \frac{\theta(t-4m_\pi^2)}{\sqrt{t}}
         + 4m_\pi \,\delta(t-4m_\pi^2) \right],
\nn
    && \Im \beta(t) = \frac{\pi}{16}\, E_\pi^2 \left[
       \Big(2 -\frac{4m_\pi^2}{t}\Big)
       \frac{\theta(t-4m_\pi^2)}{\sqrt{t}}
         - 4m_\pi \,\delta(t-4m_\pi^2) \right].
\eeqn
Accordingly, Eq.\ (\ref{Yukawa}) results in
\beqn
    4\pi r^2\alpha_L(r) &=& \frac{E_\pi^2}{2}
      \Big[ (1+z_\pi) e^{-z_\pi} + \frac{z_\pi^2}{2}\, {\rm Ei}(-z_\pi) \Big],
\nn
    4\pi r^2\beta(r) &=& \frac{E_\pi^2}{4}
      \Big[ (1-z_\pi) e^{-z_\pi} - \frac{z_\pi^2}{2}\, {\rm Ei}(-z_\pi) \Big],
\eeqn
where ${\rm Ei}(-z) = -\int_z^\infty (e^{-t}/t)\,dt$
is the exponential integral.
   The contribution of kaon loops is obtained by the substitutions
$E_\pi\to E_K$ and $m_\pi\to m_K$ as
explained in Section~\ref{sec:spatial}.

   For pions and kaons, the generalized polarizabilities of Eq.\ (\ref{betasM})
have branching points at $t=4m_\pi^2$ and $4m_K^2$ but no poles:
\beq
    \Im \beta(t) = \frac{e^2}{4\pi M}\,\frac{2\pi}{(4\pi F_\pi)^2}\,
        \left[ \Big(\frac{B_\pi}{x_\pi} + C_\pi\Big)
          \frac{\theta(x_\pi-4)}{\sqrt{x^2_\pi - 4x_\pi}}
    + C_K \frac{\theta(x_K-4)}{\sqrt{x_K^2 - 4x_K}} \right].
\eeq
Then Eq.~(\ref{Yukawa}) gives
\beqn
    4\pi r^2 \beta(r) &=& \frac{e^2}{4\pi M}\,\frac{1}{(4\pi F_\pi)^2}\,
        \Big[ 2m_\pi C_\pi  z_\pi  K_0(z_\pi)
            + 2m_K   C_K    z_K    K_0(z_K)
\nn  && \qquad {}
       + \frac{m_\pi}{2}\, B_\pi z_\pi^2 \Big( K_1(z_\pi) -
         \int_{z_\pi}^\infty K_0(x)\,dx\Big) \Big].
\eeqn


\begin{references}

\bibitem{guichon98}
  P.A.M. Guichon and M. Vanderhaeghen,
  Prog. Part. Nucl. Phys. {\bf 41}, 125 (1998). 

\bibitem{scherer99}
  S. Scherer,
  Czech. J. Phys. {\bf 49}, 1307 (1999).  

\bibitem{vanderhaeghen00}
  M. Vanderhaeghen,
  Eur. Phys. J. A {\bf 8}, 455 (2000).  


\bibitem{guichon95}
  P.A.M. Guichon, G.Q. Liu, and A.W. Thomas,
  Nucl. Phys. {\bf A 591}, 606 (1995).

\bibitem{kroll96}
  P. Kroll, M. Sch\"urmann, and P.A.M. Guichon,
  Nucl. Phys. {\bf A 598}, 435 (1996).  
\bibitem{radyushkin97}
  A.V. Radyushkin, Phys. Rev. D {\bf 56}, 5524 (1997).  
\bibitem{ji97}
  X. Ji, Phys. Rev. D {\bf 55}, 7114 (1997).  


\bibitem{Low_1954}
   F. E. Low, Phys. Rev. {\bf 96}, 1428 (1954).
\bibitem{GellMann_1954}
   M. Gell-Mann and M. L. Goldberger, Phys. Rev. {\bf 96}, 1433 (1954).
\bibitem{Scherer_96}
   S. Scherer, A. Yu. Korchin, and J. H. Koch, Phys. Rev. C {\bf 54},
   904 (1996).  

\bibitem{roche00}
  J. Roche {\it et al.},
  Phys. Rev. Lett. {\bf 85}, 708 (2000). 
\bibitem{Audit_1993}
   G. Audit {\it{et al.}}, CEBAF Report No. PR 93-050, 1993,
   http://www.jlab.org/ {\verb+~+}luminita/vcs.html.
\bibitem{Shaw_1997}
   J. Shaw {\it{et al.}}, MIT--Bates proposal No. 97-03, 1997.
\bibitem{moinester99}
  M.A. Moinester {\it et al.} (The SELEX Collaboration),
  Tel Aviv University Report No. TAUP-2568-99, hep-ex/9903039.
\bibitem{Ocherashvili00}
  A. Ocherashvili, PhD thesis, Tel Aviv University, 2000.

\bibitem{liu96}
  G.Q. Liu, A.W. Thomas, and P.A.M. Guichon,
   Austral. J. Phys. {\bf 49}, 905 (1996).  
\bibitem{pasquini98}
  B. Pasquini and G. Salm\`e,
   Phys. Rev. C {\bf 57}, 2589 (1998).  
\bibitem{pasquini00a}
  B. Pasquini, S. Scherer, and D. Drechsel, Phys. Rev. C {\bf 63}, 025205
  (2001).  

\bibitem{metz96}
  A. Metz and D. Drechsel,
    Z. Phys. A {\bf 356}, 351 (1996);
   {\it ibid.} {\bf 359}, 165 (1997). 

\bibitem{hemmert97}
  T.R. Hemmert, B.R. Holstein, G. Kn\"ochlein, and S. Scherer,
    Phys. Rev. D {\bf 55}, 2630 (1997);   
    Phys. Rev. Lett. {\bf 79}, 22 (1997).  
\bibitem{hemmert00}
  T.R. Hemmert, B.R. Holstein, G. Kn\"ochlein, and D. Drechsel,
    Phys. Rev. D {\bf 62}, 014013 (2000).  


\bibitem{kim97}
  M. Kim and D.-P. Min, hep-ph/9704381.

\bibitem{vanderhaeghen96}
  M. Vanderhaeghen, Phys. Lett. B {\bf 368}, 13 (1996).
\bibitem{korchin98}
  A.Yu. Korchin and O. Scholten,
    Phys. Rev. C {\bf 58}, 1098 (1998).

\bibitem{pasquini00b}
  B. Pasquini, D. Drechsel, M. Gorchtein, A. Metz, and M. Vanderhaeghen,
    Phys. Rev. C {\bf 62}, 052201 (2000).  
\bibitem{pasquini01}
   B. Pasquini, M. Gorchtein, D. Drechsel, A. Metz, and M. Vanderhaeghen,
   hep-ph/0102335.



\bibitem{unkmeir00}
  C. Unkmeir, S. Scherer, A.I. L'vov, and D. Drechsel,
    Phys. Rev. D {\bf 61}, 034002 (2000). 
\bibitem{fuchs00}
  T. Fuchs, B. Pasquini, C. Unkmeir, and S. Scherer,  hep-ph/0010218.


\bibitem{tarrach75}
  R. Tarrach, Nuovo Cimento A {\bf 28}, 409 (1975).
\bibitem{bardeen68}
  W.A. Bardeen and W.-K. Tung, Phys. Rev. {\bf 173}, 1423 (1968).
\bibitem{drechsel97}
  D. Drechsel, G. Kn\"ochlein, A. Metz, and S. Scherer,
  Phys. Rev. C {\bf 55}, 424 (1997).  


\bibitem{Ward_50}
   J. C. Ward, Phys. Rev. {\bf 78}, 182 (1950).
\bibitem{Fradkin_56}
   E. S. Fradkin, Sov. Phys. JETP {\bf 2}, 361 (1956).
\bibitem{Takahashi_57}
   Y. Takahashi, Nuovo Cim. {\bf 6}, 371 (1957).
\bibitem{lvov87}
   A.I. L'vov, LPI preprint \#344, Moscow (1987).
\bibitem{Chisholm_61}
   J. Chisholm, Nucl. Phys. {\bf 26}, 469 (1961).
\bibitem{Kamefuchi_61}
   S. Kamefuchi, L. O'Raifeartaigh, and A. Salam,
Nucl. Phys. {\bf 28}, 529 (1961).
\bibitem{fearing98}
  H.W. Fearing and S. Scherer,
    Few-Body Syst. {\bf 23}, 111 (1998).  

\bibitem{choudhury68}
   S.R. Choudhury and D.Z. Freedman, Phys. Rev. {\bf 168}, 1739 (1968).



\bibitem{krajcik74}
  R.A. Krajcik and L.L. Foldy, Phys. Rev. D {\bf 10}, 1777 (1974).



\bibitem{gasser84}
   J. Gasser and H. Leutwyler, Ann. Phys. (N.Y.) {\bf 158}, 142 (1984).
\bibitem{colangelo96}
  G. Colangelo, M. Finkemeier, and R. Urech,
      Phys. Rev. D {\bf 54}, 4403 (1996).  
\bibitem{bijnens98}
  J. Bijnens, G. Colangelo, and P. Talavera,
      JHEP {\bf 5}, 14 (1998);  hep-ph/9805389.

\bibitem{bernard92a}
  V. Bernard, N. Kaiser, J. Kambor, and Ulf-G. Mei{\ss}ner,
  Nucl. Phys. {\bf B 388}, 315 (1992).
\bibitem{fearing97}
  H.W. Fearing, R. Lewis, N. Mobed, and S. Scherer,
  Phys. Rev. D {\bf 56}, 1783 (1997). 
\bibitem{robilotta01}
  M.R. Robilotta, Phys. Rev. C {\bf 63}, 044004 (2001). 



\bibitem{PDG00}
  D.E. Groom {\it et al.} (Particle Data Group),
    Eur. Phys. J. C {\bf 15}, 1 (2000).



\bibitem{lvov94}
   A.I. L'vov, talk at the CEBAF--INT workshop
  ``Probing nucleon structure by real and virtual Compton scattering,"
   Crystal Mountain Resort (USA), Sep 24--30, 1994 (unpublished).

\bibitem{bernard92}
   V. Bernard, N. Kaiser, J. Kambor, and U.-G. Mei{\ss}ner,
   Phys. Rev. D {\bf 46}, R2756 (1992).
\bibitem{butler92}
   M.N. Butler and M.J. Savage,
   Phys. Lett. B {\bf 294}, 369 (1992).  

\bibitem{gasser85}
   J. Gasser and H. Leutwyler, Nucl. Phys. {\bf B250}, 465 (1985).


\bibitem{terentev72}
  M.V. Terent'ev, Yad. Fiz. {\bf 16}, 162 (1972)
       [Sov. J. Nucl. Phys. {\bf 16},  87 (1973)].


\bibitem{fuchs99}
   T. Fuchs, Diploma thesis, Johannes Gutenberg-Universit\"at, Mainz, (1999)
   (unpublished).

\bibitem{donoghue99}
  J.F. Donoghue, B.R. Holstein, and B. Borasoy,
  Phys. Rev. D {\bf 59}, 036002 (1999).  

\bibitem{lvov93}
   A.I. L'vov, Int. J. Mod. Phys. A {\bf 8}, 5267 (1993).
\bibitem{bernard93}
   V. Bernard, N. Kaiser, A. Schmidt, and U.-G. Mei{\ss}ner,
     Phys. Lett. B {\bf 319}, 269 (1993).  

\bibitem{scoccola90}
   N.N. Scoccola and W. Weise, Nucl. Phys. A {\bf 517}, 495 (1990).
\bibitem{scherer92}
   S. Scherer and P.J. Mulders,
   Nucl. Phys. {\bf A 549}, 521 (1992).
\bibitem{gobbi96}
   C. Gobbi, C.L. Schat, and N.N. Scoccola,
   Nucl. Phys. {\bf A 598}, 318 (1996).   
\bibitem{lvov97}
   A.I. L'vov, V.A. Petrun'kin, and M. Schumacher,
     Phys. Rev. C {\bf 55}, 359 (1997).


\end{references}
\end{document}